\documentclass{article}
\pdfoutput=1
\usepackage[utf8]{inputenc}
\usepackage{jcappub}
\usepackage[dvipsnames]{xcolor}
\usepackage{amssymb,amsmath}
\usepackage{subcaption}
\usepackage{graphicx}
\usepackage{multirow}
\usepackage{nicefrac}
\usepackage{url}
\usepackage{hyperref}

\newcommand{\kpar}{k_{\parallel}}
\newcommand{\kperp}{k_{\perp}}

\newcommand{\br}{\mathbf{r}}
\newcommand{\bs}{\mathbf{s}}
\newcommand{\bd}{\mathbf{d}}
\newcommand{\bq}{\mathbf{q}}
\newcommand{\bx}{\mathbf{x}}
\newcommand{\bk}{\mathbf{k}}
\newcommand{\hMpc}{\,h\,{\rm Mpc}^{-1}}
\newcommand{\Mpch}{\,h^{-1}{\rm Mpc}}
\newcommand{\HI}{{\rm HI}}
\newcommand{\rec}{{\rm rec}}

\newcommand{\rsp}[1]{{\color{black}#1}}

\author[a,b]{Chirag Modi,}
\author[a,b,c]{Martin White,}
\author[d]{An\v{z}e Slosar,}
\author[a,b]{Emanuele Castorina}

\affiliation[a]{Department of Physics, University of California, Berkeley, CA 94720}
\affiliation[b]{Berkeley Center for Cosmological Physics, Berkeley, CA 94720}
\affiliation[c]{Department of Astronomy, University of California, Berkeley, CA 94720}
\affiliation[d]{Department of Physics, Brookhaven National Laboratory, Upton, NY 11973}

\emailAdd{modichirag@berkeley.edu}
\emailAdd{mwhite@berkeley.edu}
\emailAdd{anze@bnl.gov}
\emailAdd{ecastorina@berkeley.edu}

\title{Reconstructing large-scale structure with neutral hydrogen surveys}
\date{December 2018}

\abstract{Upcoming 21-cm intensity surveys will use the hyperfine transition in emission to map out neutral hydrogen in large volumes of the universe. Unfortunately, large spatial scales are completely contaminated with  spectrally smooth astrophysical foregrounds which are orders of magnitude brighter than the signal. This contamination also leaks into smaller radial and angular modes to form a foreground wedge, further limiting the usefulness of 21-cm observations for different science cases, especially cross-correlations with tracers that have wide kernels in the radial direction. In this paper, we investigate reconstructing these modes within a forward modeling framework. Starting with an initial density field, a suitable bias parameterization and non-linear dynamics to model the observed 21-cm field, our reconstruction proceeds by \rsp{combining} the likelihood of a forward simulation to match the observations (under given modeling error and a data noise model) \rsp{with the Gaussian prior on initial conditions and maximizing the obtained posterior}. For redshifts $z=2$ and $4$, we are able to reconstruct 21cm field with cross correlation, $r_c > 0.8$ on all scales for both our optimistic and pessimistic assumptions about foreground contamination and for different levels of thermal noise.  The performance deteriorates slightly at $z=6$.  The large-scale line-of-sight modes are reconstructed almost perfectly. We demonstrate how our method also provides a technique for density field reconstruction for baryon acoustic oscillations, outperforming standard methods on all scales. We also describe how our reconstructed field can provide superb clustering redshift estimation at high redshifts, where it is otherwise extremely difficult to obtain dense spectroscopic samples, as well as open up a wealth of cross-correlation opportunities with projected fields (e.g.\ lensing) which are restricted to modes transverse to the line of sight.
}

\begin{document}
\maketitle
\flushbottom

\section{Introduction}

The coming decade will see transformative dark energy science done by both ground-based surveys (DESI \cite{DESI}, LSST \cite{LSST}) and space-based missions (Euclid \cite{EUCLID18}, WFIRST \cite{WFIRST18}). We expect that large swaths of the universe will be sampled to close to the sample variance limit on very large scales using galaxies as tracers of density field as well as sources used to measure the weak gravitational lensing shear or cosmic backlights illuminating the cosmic hydrogen for Lyman-$\alpha$ forest studies. At the same time, these fields will offer a multitude of cross-correlation opportunities, through which more and more robust science will be derived. 

Looking beyond the current decade, several experimental options are being considered that will allow us to continue on the path of mapping ever increasing volume of the Universe. Photometric experiments, such as LSST, will likely be systematics limited by the quality of the photometric redshifts. Spectroscopic instruments are more attractive, especially using LSST as the targetting survey, but will require major investments in a dedicated new telescope and more aggressive spectrograph multiplexing to be truly interesting when compared to DESI. Going at higher redshift, traditional galaxy spectroscopy in optical becomes ever more difficult, since the sources are sparser, fainter and further redshifted. On the other hand,  turning to radio and relying on the 21-cm signal from neutral hydrogen could offer a cost-effective way of reaching deep sampling of the universe, especially in the redshifts range $2 \lesssim z \lesssim 6$, which remains largely unexplored on cosmological scales.

A number of upcoming (CHIME \cite{CHIME}, HIRAX \cite{HIRAX}, Tianlai \cite{Tianlai}, SKA\cite{SKACosmo}) or planned (PUMA, the proposed Stage {\sc ii} experiment \cite{CVDE-21cm}) interferometric instruments will use the 21-cm signal to probe this volume of the universe.  In the post-reionization era most of the hydrogen in the Universe is ionized, and the 21-cm signal comes from self-shielded regions largely within halos.  Such `intensity mapping' surveys therefore measure the unresolved emission from halos tracing the cosmic web in redshift space. 

The 21-cm observations suffer from one fundamental problem, namely that it is completely insensitive to modes which vary slowly along the line of sight (that is low $\kpar$ modes), because these are perfectly degenerate with foregrounds which are orders of magnitude brighter that the signal \cite{Shaw14,Shaw15,Pober15,Byrne18}. This effect alone severely limits the usefulness of 21-cm observations for cross-correlations with tracers that have wide kernels in the radial directions. This is in particular true for cross-correlations with the cosmic microwave background (CMB) lensing reconstructions and shear field measured by the photometric galaxy surveys, such as that coming from LSST. Cross-correlations with CMB would allow us to use the 21-cm observations and CMB observations in conjunction to put the strongest limits on modified gravity by measuring growth. Cross-correlations with spectroscopic galaxies would allow us to characterize the source redshift of galaxy samples used for weak-lensing measurements, one of the main optical weak lensing systematics. Moreover, the foreground wedge (discussed below) can render further regions of the $k$-space impotent for cosmological analysis \cite{Datta10,Morales12,Parsons12,Shaw14,Shaw15,Liu14,Pober15,SeoHir16,Cohn16,Byrne18}. The foreground wedge is not as fundamental as the low $\kpar$ foreground contamination, because it only results as a consequence of an imperfect instrument calibration rather than being a fundamental astrophysical bane.  Nevertheless, it is a data cut that will likely be necessary at least in the first iterations of data analysis from the upcoming surveys.

Can this lost information be recovered? The answer is yes, although the extend to which this is possible depends on both the resolution and noise properties of the instrument. Imagine the following thought experiment. At infinite resolution in real space (and ignoring the thermal broadening of the signal for the moment), the underlying radio intensity is composed of individual objects, that appear as distinct peaks. A high pass filter in $k$ space will not fundamentally alter one's ability to count these objects.  While the filtering might introduce `wings' in individual profiles, the peaks in the density fields are still there. In this case, we can recover the lost large-scale modes perfectly. A somewhat different way to look at the same physics is to realize that the non-linear mode coupling propagates information from large-scale to small-scale modes, while the initial conditions of the small-scale modes are forgotten. Since the total number of modes scales as $k^2\,\delta k$, there are always many more small-scale modes than large-scale modes and in a sense the system is over-constrained if one wants to recover large-scale modes for which there is no direct measurements. Non-linear evolution erases the primordial phase information on small scales and encodes the primoridal large-scale field over the entire $k$-space volume. So it is clear that the process of backing out the large-scale information from the small scale is possible, at least in the limit of sufficiently low noise and sufficiently high resolution.

In this paper we approach this problem by means of forward modelling the final density field. Starting with an initial density field and a suitable bias parameterization, we reformulate the problem of recovering the large scales as a problem of non-linear inversion. In the past few years, several reconstruction methods have been developed to solve this \cite{Jasche13,Seljak17,Schmittfull17,Modi18,Schmidt18,Schmittfull19}. The solution of this non-linear problem is the linear, 3D field that evolves under the given forward model to result in observed final field. Thus one not only recovers the linear large scales where the information was gone, but also automatically performs an optimal reconstruction of the linear field \emph{on scales where we had non-linear information to start with}. This is often the product that one ultimately desires. It allows not only optimal BAO measurement for the 21-cm survey, but also increases the scale over which one can model cross-correlations with other tracers, thus enabling cross-correlations with CMB lensing reconstruction and photometric galaxy samples.

The full implications of this reconstruction exceed the scope of this paper. Instead we focus on the basic questions: 
i) Does the forward modeling approach to reconstruction work at all in the case where we loose linear modes to foregrounds? ii) What is the complexity of forward model required? iii) How does the result depend on the noise and angular resolution of the experiment? iv) How does the performance vary with scale, direction, redshift and real vs.\ redshift space data? v) What are the gains one expects for different science objectives such as BAO reconstruction, cross-correlations with CMB and cross-correlations with different LSS surveys such as LSST. 

The outline of the paper is as follows.  We begin by discussing the observational constraints we are likely to face in \S\ref{sec:instruments} and the simulation suite we will be using as mock data to demonstrate our reconstruction algorithm in \S\ref{sec:sims}. Next, we review our forward model and the method we use to reconstruct the field from the observations in \S\ref{sec:recon} (building upon refs.~\cite{Seljak17,Modi18}). In \S\ref{sec:results} we show the results for our forward model on the `observed' 21-cm data, as well as gauge the performance of our reconstruction algorithm on different metrics for multiple experimental setups.  In \S\ref{sec:implications} we show the improvements expected by using our reconstructed field for different science objectives such as BAO reconstruction, photometric redshift estimation and CMB cross correlations.  We conclude in \S\ref{sec:conclusions}.

\section{Observational constraints}
\label{sec:instruments}

The instruments of interest for this investigation are interferometers measuring the redshifted 21-cm line.  Such instruments work in the Fourier domain with the correlation between every pair of feeds, $i$ and $j$, measuring the Fourier transform of the sky emission times the primary beam at a wavenumber, $k_\perp = 2\pi \vec{u}_{ij}/ \chi(z)$, set by the spacing of the two feeds in units of the observing wavelength ($\vec{u}_{ij}$) \cite{TMS17}.
The visibility noise is inversely proportional to the number (density) of such baselines \cite{ZalFurHer04,McQuinn06,Seo2010,Bull2015,SeoHir16,Cohn16,Wol17,Alonso17,White17,Obuljen18,CVDE-21cm,Chen19}.
Where necessary, we take the noise parameters from Refs.~\cite{CVDE-21cm,Chen19}.
We shall investigate how our reconstruction depends upon the noise (thermal plus shot noise) and the $k$-space sampling of the instrument.

Radio foregrounds, primarily free-free and synchrotron emission from the Galaxy and unresolved point sources, are several orders of magnitude brighter than the signal of interest and present a major problem for 21-cm measurements \cite{Furlanetto06,Shaw14,Pober15}. However, due to their emission mechanisms, they are intrinsically very spectrally smooth and this is the property that allows them to be separated from the signal of interest which varies along the line of sight due to variation in underlying cosmic density field along the line of sight. This separation naturally becomes increasingly difficult as we seek to recover very low $k_\parallel$ modes, i.e.\ modes close to transverse to the line of sight.  The precise value below which recovery becomes impossible is currently unknown (see Refs.~\cite{Shaw14,Shaw15,Pober15,Byrne18} for a range of opinions). To be conservative, we will assume that we lose all the modes below $k_\parallel=0.03 \hMpc$, however we will also study how sensitive are our results to this cut-off value.

In addition to low $k_\parallel$, non-idealities in the instrument lead to leakage of foreground information into higher $k_\parallel$ modes. This arises because, for a single baseline, a monochromatic source (i.e.\ a bright foreground) at non-zero path-length difference is perfectly degenerate with signal at zero path-length difference (i.e.\ zenith for transiting arrays) but appropriately non-flat spectrum, such as that arising from 21-cm fluctuations. This is usually phrased in terms of a foreground ``wedge'' which renders modes with low $k_\parallel/k_\perp$ unusable \cite{Datta10,Morales12,Parsons12,Shaw14,Shaw15,Liu14,Pober15,SeoHir16,Cohn16,Byrne18}. Due to the variation of Hubble parameter with redshift the wedge becomes progressively larger at higher redshift. Information in this wedge is not irretrievably lost, because using multiple baselines can break the degeneracy, but it requires progressively better phase calibration the deeper into the wedge one pushes.  The better the instrument can be calibrated and characterized the smaller the impact of the wedge. The most pessimistic wedge assumption is that all sources to the horizon contribute to the contamination -- will we not consider this case as it makes a 21-cm survey largely ineffective for large-scale structure. The most optimistic assumption is that the wedge has been subtracted perfectly.  We regard this as unrealistic. A middle-of-the-road assumption is that the wedge matters to an angle measured by primary field of view. We take our `optimistic' choice to be the `primary beam’ wedge defined as $\theta_w = 1.22\lambda/2D_e$, where $D_e$ is the effective diameter of the dish after factoring in the aperture efficiency ($\eta_\alpha = 0.7$) and the factor of two in the denominator gives an approximate conversion between the first null of the Airy disk and its full Width at half maximum (FWHM). We shall contrast this with the `pessimistic'  case $\theta_w = 3\times 1.22\lambda/2D_e$.
 
\begin{figure}
    \centering
    \resizebox{\columnwidth}{!}{\includegraphics{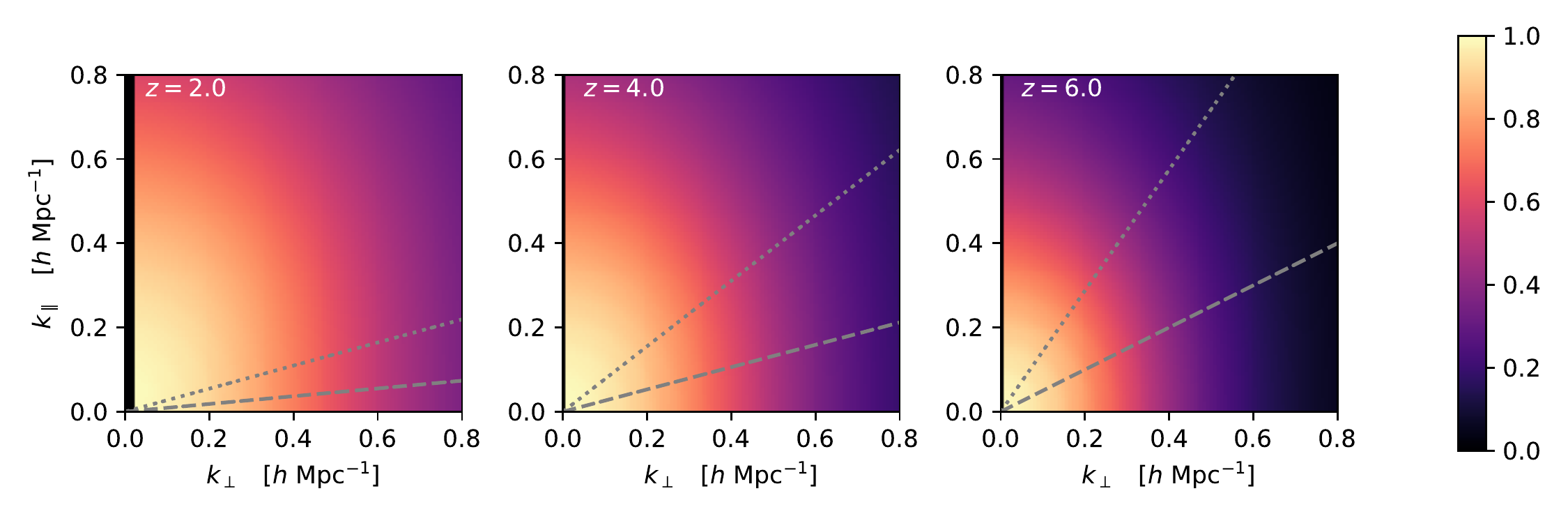}}
    \caption{The fraction of the total power that is signal, $S/(S+N)$, vs.~$k_\perp$ and $k_\parallel$ for a $5\,$yr PUMA-like survey at $z=2$ (left), 4 (middle) and 6 (right).  The dotted and dashed lines in each panel show a pessimistic and optimistic forecast for the foreground wedge (see text).  The loss of low $k_\perp$ modes, most visible at $z=2$, is due to the constraint that dishes must be further apart than their diameters, leading to a minimum baseline length.}
\label{fig:knoise}
\end{figure}

Fig.~\ref{fig:knoise} shows this information in graphical form, following refs.~\cite{Chen19,HiddenValley19}.  The color scale shows the fraction of the total power which is signal, as a function of $k_\perp$ and $k_\parallel$.  The modes lost to foregrounds in the wedge are illustrated by the gray dashed (optimistic) and dotted (pessimistic) lines.  Modes below and to the right of these lines would be contaminated by foregrounds.  In addition we expect to lose modes with $k_\parallel<k_\parallel^{\rm min}$ where $k_\parallel^{\rm min}=0.01-0.1\,h\,{\rm Mpc}^{-1}$.  At $z=2$ and low $k$ the signal dominates, at intermediate $k$ the shot-noise starts to become important and at high $k_\parallel$ the thermal noise from the instrument dominates.

\section{Data: Hidden Valley simulations}
\label{sec:sims}

To test the efficacy of our method we make use of the Hidden Valley\footnote{http://cyril.astro.berkeley.edu/HiddenValley} simulations \cite{HiddenValley19}, a set of trillion-particle N-body simulations in gigaparsec volumes aimed at intensity mapping science.  Our workhorse simulation will be {\tt HV10240/R}, which evolved $10240^3$ particles in a periodic, $1024\,h^{-1}$Mpc box from Gaussian initial conditions using a particle-mesh code \cite{FengChuEtAl16} with a $20480^3$ force mesh. At this resolution, one is able to resolve halos down to $M\sim 10^9\,h^{-1}{\rm M}_\odot$, which host the majority ($>95\%$) of the cosmological HI signal, while the volume allows robust measurement of observables such as baryon acoustic oscillations.

The halos in the simulation were assigned neutral hydrogen with a semi-analytic recipe, outlined in more detail in ref.~\cite{HiddenValley19}.  We make use of their fiducial model, `Model A', at $z=2$, $4$ and $z=6$. Briefly, this model populates halos with centrals and satellites following a halo occupation description and these galaxies are then assigned HI mass following a $M_h-M_{\rm HI}$ relation. Our mock HI data lives in redshift space and captures the small scale non-linear redshift space distortion effects due to satellite motion.
We also caution the reader that \rsp{while we have made use of a particular semi-analytic model, 
the manner in which HI traces the matter field at high $z$ is currently poorly constrained 
observationally.  While our model is consistent with our best current knowledge, the particular values of various modeling parameters we have assumed may not be correct.
However as long as this field can be modeled with a flexible bias framework and the scatter between the HI mass and halo mass (or underlying dark matter density) is similar to other tracers, such as stellar mass,} we believe our \rsp{qualitative} results do not depend critically upon these details.

\rsp{
For our analysis, we will use two Cartesian meshes (discussed further in Section \ref{sec:annealing})
with 256 and 512 cells along each dimension, which have resolutions of $4$ and $2\Mpch$ respectively.
To generate the HI data, we desposit the galaxies, weighted by their HI mass, on these meshes with a cloud-in-cloud (CIC) interpolation scheme and then estimate the HI overdensity field ($\delta_{\rm HI}$) in the usual way.
Similarly, to generate the final, Eulerian matter field at the redshifts of interest, we use a $4\%$ subsampled snapshot of the particle positions and paint them with the same CIC scheme (and equal weights per particle). 
The initial, Lagrangian field ($\delta_L$), is generated on the meshes using the same initial seed (hence same initial conditions) as the simulation itself.

Once we generate the `clean HI data' on a mesh we need to simulate the foreground wedge and thermal noise to construct a mock observed data for the purpose of reconstruction.
To include the foreground wedge, $w$, we simply omit from our calculations all of the modes that are within the wedge, i.e.\ below a certain cut-off $k_\parallel/k_\perp$, as shown in Fig.~\ref{fig:knoise}. 
We simulate thermal noise by drawing a zero-mean, Gaussian realization, $n(\bk)$, from the 2D thermal noise power spectrum, $P_{\rm th}(k, \mu)$.
Then, we corrupt the `clean' data by adding this noise realization and use it as the `observed' data for reconstruction
\begin{equation}
  \delta_{\rm HI}^n(\bk) = \delta_{\rm HI}(\bk) + n(\bk)
  \label{eq:addnoise}
\end{equation}
}

\section{Reconstruction Method}
\label{sec:recon}

There have been many approaches to determining the density field from noisy or incomplete observations in cosmology, from early work like refs.~\cite{Peebles89,Bertschinger89,Rybicki92,Nusser92,Dekel94} through ref.~\cite{Narayanan99} and references therein.  Forward model reconstructions similar to our approach have also been developed previously and applied in other contexts \cite{Jasche10,Jasche13,Kitaura13,Wang14,Jasche15,Wang16}.
Other authors have also investigated how one could reconstruct the long-wavelength modes which are lost to foregrounds in intensity mapping \cite{Karacay19,Others}.  Ref.~\cite{Karacay19}, in particular, investigates a very similar problem with a related approach. 

Our method to reconstruct the long-wavelength modes puts in practice the intuition developed in the introduction, that the non-linear evolution of matter under gravity propagates information from large scales in the initial conditions to the small scales in the final matter field. We reconstruct the initial density field by optimizing its posterior, conditioned on the observed data (HI density in $k$-space), assuming Gaussian initial conditions. Evolving this initial field allows us to reconstruct the observed data on all scales, thus reconstructing the sought-after long wavelength modes. To solve the problem of reconstruction of initial conditions, we follow the approach developed by refs.~\cite{Seljak17,Modi18}. In this first application of the method we will hold the cosmology fixed, though in future one could imagine jointly fitting the cosmology and the long wavelength modes.
\rsp{Varying cosmology and hence the prior power spectrum is especially important when quantifying uncertainities on the reconstructed fields but that is beyond the scope of this work and we defer it for future.}

\rsp{In this section, we begin by outlining our forward model with the focus on the bias model. Next, we use this forward model to setup the likelihood function for reconstruction and modify it to account for the additional noise and foreground wedge present in the observed data. Lastly, we also outline annealing schemes that we develop to assist convergence of our algorithm and hence improve our reconstructions.}

\subsection{Forward Model}
\label{sec:biasmodel}

To use the framework of refs.~\cite{Seljak17,Modi18}, we require a forward model [$\mathcal{F}(\bs)$] to connect the observed data $(\bd)$ with the Gaussian initial conditions (ICs; $\bs$) in a differentiable manner.  This forward model is typically composed of two parts - non-linear dynamics to evolve dark-matter field from the Gaussian ICs to the Eulerian space and a mapping from the underlying matter field to the observed tracers \rsp{i.e.\ a bias model}. 
We try two different models to shift the particles from Lagrangian space to their Eulerian space: a simple Zeldovich displacement \cite{Zel70} and full N-body evolution \cite{FengChuEtAl16}, albeit at relatively low resolution.

For the mapping from the matter field to the tracers we use a Lagrangian bias model (as for example in refs.~\cite{Matsubara08a,Matsubara08b,Carlson13,White14,Vlah16,Modi16,Schmittfull19}) including terms up to quadratic order.
At this order, the Lagrangian fields are $\delta_L(\bq)$, $\delta_L^2(\bq)$ and $s^2_L(\bq) \equiv \sum_{ij}s_{ij}^2(\bq)$ 
the scalar shear field where $s_{ij}^2(\bq) = (\partial_i \partial_j\partial^{-2} - [1/3]\delta_{ij}^{D}) \delta_L(\bq)$
-- we discuss derivative bias below.
We use these fields from the ICs of the simulation itself (\rsp{as described in Section \ref{sec:sims}}),
but evolved to $z=0$ using linear theory (due to this evolution, care must be taken in interpreting our bias parameters). 
Since $\delta_L^2(\bq)$ and $s^2(\bq)$ are correlated, we define a new field $g_L^2(\bq) = \delta_L^2(\bq) - s^2(\bq)$ which does not correlate with $\delta_L^2(\bq)$ on large scales and use this instead of shear field.  In addition, we subtract the zero-lag terms to make these fields have zero mean.

\rsp{Then, to generate our model field as a combination of these Lagrangian fields, 
we use the approach developed by ref.~\cite{Schmittfull19}, which is itself an implementation
of the ideas of ref.~\cite{Matsubara08b} (e.g.~Eq.~8) and its extensions \cite{Vlah16}. 
Specifically we match the model with the observations at the level of the field instead of only matching the two-point functions. 
We use the $512^3$ particle mesh for these fields, with particle/mesh spacing $2\,h^{-1}$Mpc for our box of $1024\Mpch$ (and no further smoothing).
Each particle is `shifted' to its Eulerian position, $\bx$, with the non-linear dynamics of choice and then binned onto the grid with cloud-in-cloud (CIC) interpolation and weight
\begin{equation}
 {\rm weight} = 1 + b_1\delta_L(\bq) + b_2 (\delta^2_L(\bq) - \langle \delta^2_L(\bq)\rangle) + b_g \left(g_L(\bq) - \left\langle g_L(\bq) \right\rangle\right)
\end{equation}
Note the contributions of $\delta_L$, $\delta_L^2$ and $g_L^2$ assigned to each particle are based on its initial location ($\bq$).
This procedure is equivalent to building separate fields with each particle weighted by $1$, $\delta_L$, $\delta_L^2$ and $g_L$ and then taking the linear combination of those fields after shifting, which is how the model was actually implemented.
Thus our modeled tracer field is:
\begin{equation}
    \delta_{\rm HI}^b(\bx) = \delta_{[1]}(\bx) + b_1\delta_{[\delta_L]}(\bx) + b_2\delta_{[\delta^2_L]}(\bx)  + b_g \delta_{[g_L]}(\bx) \qquad .
\label{eq:biasmodel1}
\end{equation}
where $\delta_{[W]}(\bx)$ refers to the field generated by shifting the particles weighted with `$W$' field.
}

To fit for the bias parameters, we minimize the mean square model error between the data and the model fields which is equivalent to minimizing the error power spectrum of the residuals, $\br(k) = \delta_{\rm HI}^b(\bk) - \delta_{\rm HI}(\bk)$, in Fourier space: 
\begin{equation}
    P_{\rm err}(k) = \frac{1}{N_{\rm modes}(k)}\sum_{\bk, |\bk|\sim k}\left|\delta_{\rm HI}^b(\bk) - \delta_{\rm HI}(\bk)\right|^2
\label{eq:error}
\end{equation}
where the sum is over half of the $\mathbf{k}$ plane since $\delta^\star(\mathbf{k})=\delta(-\mathbf{k})$ for the Fourier transform of a real field, and the `data' correspond to our `clean' field with no noise at this stage.

In principle the bias parameters can be made scale dependent, $b(k)$, and treated as transfer functions \cite{Schmittfull19}. To get these transfer functions, we simply minimize Eq.~\ref{eq:error} for every $k$-bin independently.  However we will find that the best-fit parameters are scale independent to a very good degree (see below).  Thus to minimize the number of fitted parameters we use scalar bias parameters and fit for them by minimizing the error power spectrum only on large scales, $k < 0.3 \hMpc$. We will show below that the fit is quite insensitive to this $k$-range (chosen reasonably) used for fitting the bias parameters.

Traditionally, bias parameters are defined such that the bias model field matches the observations in real space. However the observations of 21-cm surveys are done in redshift space. To model these observations, we `shift' the Lagrangian field directly into redshift space and minimize the error power spectrum (Eq.~\ref{eq:error}) directly in redshift space instead of real space.

Finally, we shall assume throughout that the cosmological parameters will be well known, and in fact use the values assumed in the simulation.  In principle one could iterate over cosmological parameters, but the 21-cm data themselves (and external data) will provide us with extremely tight constraints on cosmological parameters even before reconstruction \cite{CVDE-21cm}.

\subsection{Reconstruction}
\label{sec:recon2}

Once the bias parameters are fixed the procedure above provides a differentiable `forward model', $\mathcal{F}(\bs) (= \delta_{\rm HI}^b(\bx)) $, from the Gaussian initial conditions ($\mathbf{s(\bk)}$) to the observations, $\bd(\bk) = \delta_{HI}(\bk)$.  To reconstruct the initial conditions we need a likelihood model for the data. Since the error power spectrum \rsp{was minimized to fit for the model bias parameters, it measures the remaining} disagreement between the observations and our model and hence provides a natural likelihood function. \rsp{ When using this form of likelihood, we have made the assumption that the residuals in Fourier space between our model and data are drawn from a diagonal Gaussian in the Fourier space with variance given by error power spectrum. The diagonal assumption is valid due to the translational invariance of both the data and the model. The Gaussian assumption is motivated on large scales by the fact that the dynamics are linear, and on small scales by central limit theorem when averaged over large number of modes on these scales. Moreover, while this likelihood might not be completely accurate or optimal on all scales, it does provide us well motivated and simple (negative) loss function, maximizing which should reconstruct the data. This is sufficient for our purpose here, where we are only interested in a point estimate of the reconstructed field.} Thus we can write down the negative log-likelihood:
\begin{equation}
    \mathcal{L} = \frac{1}{2}\chi^2 =  \sum_k  \frac{1}{N_{\rm modes}(k)}\sum_{\bk, |\bk|\sim k} \frac{|\delta_{\rm HI}^b(\bk) - \delta_{\rm HI}(\bk)|^2}{P_{\rm err}(k)}
    \label{eq:likelihood}
\end{equation}
where the sum is over half of the $\mathbf{k}$ plane since $\delta^\star(\mathbf{k})=\delta(-\mathbf{k})$ for the Fourier transform of a real field.
This is combined with the prior over the initial modes, which are assumed to be Gaussian and uncorrelated in the Fourier space. Hence the negative log-likelihood for the Gaussian prior can be combined with the negative log-likelihood of the data to get the posterior
\begin{equation}
    \mathcal{P} =   \sum_k \frac{1}{N_{\rm modes}(k)}\sum_{\bk, |\bk|\sim k} \Big( \frac{|\delta_{\rm HI}^b(\bk) - \delta_{\rm HI}(\bk)|^2}{P_{\rm err}(k)} +  \frac{|\bs(\bk))|^2}{P_{\rm s}(k)} \Big)
    \label{eq:posterior}
\end{equation}
where $P_{\rm s}$ is the initial prior power spectrum. To reconstruct the initial modes $\mathbf{s}$, we minimize this posterior with respect to them using L-BFGS\footnote{https://en.wikipedia.org/wiki/Limited-memory\_BFGS} \cite{nocedal06}, as in refs.~\cite{Seljak17,Modi18}, to get a maximum-a-posteriori (MAP) estimate.

\rsp{We note that while, in principle, one would measure the (modeling) error power spectrum as an average of different simulations, due to the computational requirements inherent in simulating the HI field we are forced to use the same single simulation to fit for the bias parameters and measure error spectra, and then use it as mock data for reconstruction.
This could potentially lead to overfitting, improving the reconstruction by underestimating the error spectra and ignoring cosmic variance.
To check this, we estimate the error power spectra for modeling the halo mass field on a set of (cheaper) simulations with smaller boxes and poorer resolution, where the problem of cosmic variance if anything should be worse.
We find that using the bias parameters fit for a one simulation and estimating the error spectra on other simulations, the error power spectrum varies.  However its ratio with the error power spectrum for the `fit' simulation is not consistently greater than 1, as one would have expected in case of overfitting but has a distribution around 1 on all scales of interest.
Hence while one would need to quantify the distribution of this error spectra to get uncertainties on the reconstructed field, we do not find any evidence that using the same simulation as mock data and to estimate error spectra leads to any overfitting or an artificially good reconstruction.}

\subsection{Noise}
\label{sec:noise}

So far we have ignored the presence of noise and the foreground wedge in our data.  While shot noise is included automatically if we use the HI realization in the simulations, we must handle foregrounds explicitly.  Due to the foreground wedge, $w$, we loose all the information in the modes below a certain cut-off $k_\parallel/k_\perp$, as shown in Fig.~\ref{fig:knoise}.  To take this into account in our reconstruction, we simply drop these modes from our likelihood term.

In addition to the wedge, 21-cm surveys also suffer from thermal noise that dominates on small scales and has angular dependence with respect to the line of sight. 
\rsp{As outlined previously in Eq. \ref{eq:addnoise}, this is incorporated by drawing a Gaussian noise realization, $n(\bk)$, with zero mean and noise power spectrum, $P_{\rm th}(k,\mu)$ that is then used to added to our simulated data ($\delta_{\rm HI}$) to generate our mock data  $\delta_{\rm HI}^n$.}
\rsp{Including both effects our final posterior is :}
\begin{equation}
    \mathcal{P}_w =  \sum_k \frac{1}{\rm N_{modes}(k)} \left( \sum_{\substack{\bk, |\bk|\sim k, \\ {\bk \not\in w}}}  \frac{|\delta_{\rm HI}^b(\bk) - \delta_{\rm HI}^n(\bk)|^2}{P_{\rm err}(k,\mu)} +  \sum_{\bk, |\bk|\sim k}  \frac{|\bs(\bk))|^2}{P_{\rm s}(k)}  \right)
\end{equation}
\rsp{The error power spectrum, $P_{\rm err}$, is now a combination of the modeling error (as before) and the noise power spectrum. This changes the amplitude of $P_{\rm err}$, especially on small scales, and also introduces an angular dependence.} We have indicated this by the additional $\mu$ dependence in $P_{\rm err}$ in the likelihood term of $\mathcal{P}_w$.  Note the data automatically include shot-noise, since we have a single realization of the halo field in the simulation.

\subsection{Annealing}
\label{sec:annealing}

Reconstructing the initial modes by minimizing Eq.~\ref{eq:posterior} is an optimization problem in a high dimensional space with both the number of underlying features (initial modes) and the number of data points (grid cells) being in millions. Despite using gradient and approximate Hessian information, it is a hard problem to solve.  Since we are  aware of the underlying physics driving our model, as well as its performance, we use our domain knowledge to assist the convergence of the optimizer by modifying the loss function over iterations rather than simply brute-forcing the optimization with the vanilla loss function. A more detailed discussion on these schemes is provided in refs.~\cite{Seljak17, Feng18, Modi18}. Here we briefly summarize the two annealing schemes that we use to improve our performance

\begin{itemize}
    \item Residual smoothing : Both the dynamics and the bias model are more linear on large scales than smaller scales. Hence the posterior surface is more convex on these scales and convergence is easier. However since the number of modes scales as $k^3$, the large scale modes are a small fraction of the total and are harder for the optimizer to reconstruct in practice. To mitigate this we smooth the residual term of the loss function on small scales with a Gaussian kernel. Thus on these scales, the prior pulls down the small scale power to zero and we force the optimizer to get the large scales correct first.
    
    \item Upsampling: To minimize the cost of reconstruction, we begin our optimization on a low-resolution grid and reconstruct all the modes following the residual smoothing. Upon convergence, we upsample the converged initial field to a higher resolution grid and paint our HI data on this grid as well. Since the higher resolution has information to smaller scales, this allows us to leverage these scales to improve our reconstruction. Further, since the largest scales have already converged on the lower resolution, they remain mostly unchanged and we do not need to repeat the residual smoothing on all the scales for this higher resolution.
\end{itemize}

\section{Results}
\label{sec:results}

We present the results for our reconstruction in the section. Our primary metrics to gauge the performance of our model as well as reconstruction are the cross correlation function, $r_{cc}(k)$, and transfer function, $T_f(k)$, defined as
\begin{equation}
    r_{cc}(k) = \frac{P_{XY}(k)}{\sqrt{P_{X}(k) P_{Y}(k)}} \qquad , \qquad
    T_f(k) = \sqrt{\frac{P_{Y}(k)}{P_{X}(k)}}  \quad ,
\label{eq:rt-def}
\end{equation}
and the error power spectrum, $P_{\rm err}$, defined in Eq.~\ref{eq:error}.
These metrics will always be defined between either the model or the reconstructed fields as $Y$ and the corresponding true field as $X$ unless explicitly specified otherwise.

\newcommand{\vS}{{\rm S}}
\newcommand{\vN}{{\rm N}}
To gain some intuition for these functions it may be helpful to recall the results for the linear case.  For Gaussian signal, $\mathbf{s}$, with covariance $\vS$, and \rsp{Gaussian} noise, $\mathbf{n}$, with covariance $\vN$ and a data vector $\mathbf{d}=\mathbf{s}+\mathbf{n}$, the posterior is given by $P(s|d) \propto P(d|s) P(s)$ which is the product of two Gaussians.  The MAP solution is given by the  well-known Wiener filter\footnote{https://en.wikipedia.org/wiki/Wiener\_filter} \rsp{\cite{wiener64}}
\begin{equation}
  \tilde{s}= \vS  \left( \vS + \vN \right)^{-1} \mathbf{d} \equiv W\mathbf{d} \qquad .
\end{equation}
For stationary problems both $\vS$ and $\vN$ are diagonal in Fourier space. In this simple case $r_{cc}(k)=T_f(k)=W^{1/2}$.  In the limit of very small noise, $P_N/P_S\ll 1$, the measurements are a faithtful representation of the true field and $W^2=r_{cc}=T_f\simeq 1$. In the limit of very large noise, $P_N/P_s\gg 1$, the filter becomes prior dominated and the most-likely value of $\mathbf{s}$ is zero.  In this limit $W^2=r_{cc}=T_f\rightarrow 0$.

In our case, there is a non-linear transformation at the heart of the model, i.e.\ $\mathbf{d} = \mathcal{F}(\bs) + \mathbf{n}$. Therefore the Wiener filter is no longer the solution, but much of the same intuition applies. In particular $r_{cc}$ measures how faithfully the reconstructed map describes 
the input map, up to rescalings of the output map amplitude. The transfer function, on the other hand, tells us about the amplitude of the output map as a function of scale, with $r\,T_f = P_{XY}/P_X$.

\subsection{Bias model}

\begin{table}[]
    \centering
    \begin{tabular}{c|cccc||cccc}
            & \multicolumn{4}{c||}{Model A} & \multicolumn{4}{c}{Model B} \\
        $z$ & $b_1$ & $b_2$ & $b_g$ & $T_f(k)$      & $b_1$ & $b_2$ & $b_g$ & $T_f(k)$  \\ \hline
          2 &  0.528 & 0.006 & -0.023    & $0.98-0.197\,k^2$ &  0.55 & -0.025 & -0.012    & $0.97-0.222\,k^2$  \\
          4 &  0.434 &  0.046 & -0.011   & $0.993-0.110\,k^2$ &  0.446 &  0.102 & -0.012    & $0.99-0.151\,k^2$ \\
          6 &  0.399 &  0.094 & -0.009  & $0.995-0.112\,k^2$ &  0.378 &  0.16 & -0.011  & $0.984-0.187\,k^2$
    \end{tabular}
    \caption{The best-fit bias parameters (and transfer function) to the HiddenValley simulation HI fields for models A and B of ref.~\cite{HiddenValley19} at $z=2$, 4 and 6.  The bias parameters are defined on a $2\,h^{-1}$Mpc grid (see text).}
    \label{tab:biasparams}
\end{table}

\begin{figure}
    \centering
    \resizebox{1\columnwidth}{!}{\includegraphics{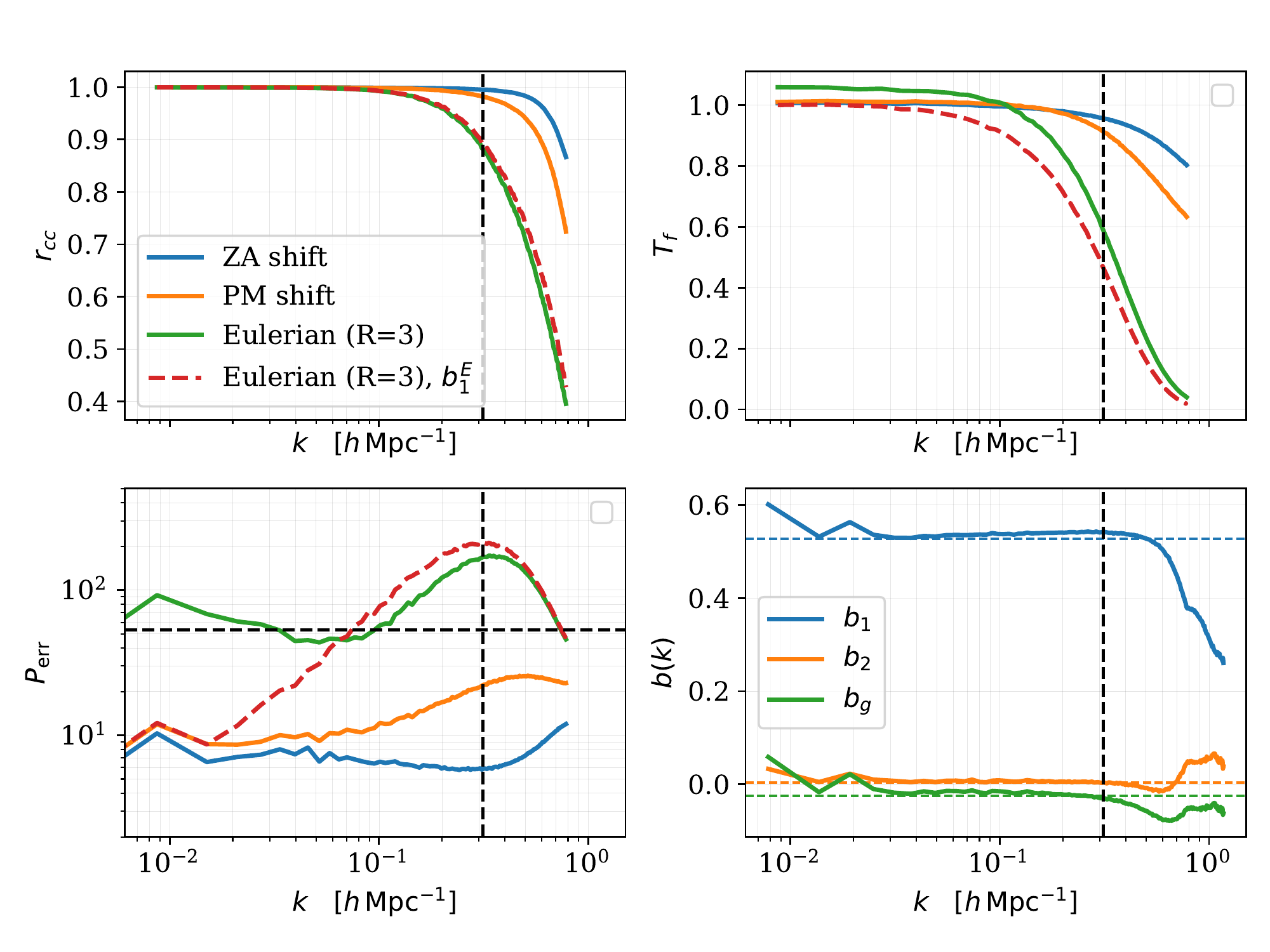}}
     \caption{Comparison of different bias models through the real-space cross-correlation (top left) and transfer function (top right) at $z=2$. Blue and orange lines correspond to the Lagrangian bias model with Zeldovich and N-body dynamics respectively. The green line corresponds to an Eulerian bias model through quadratic order, while the dashed red line is a simple linear bias model in Eulerian space. In both cases the Eulerian fields are smoothed with a Gaussian kernel of $3\Mpch$. The vertical dashed-black line corresponds to the $k_{\rm max}$ upto which the the error is minimized. (Bottom left) Comparison of different bias models at the level of the error power spectrum. The dashed black line is the `Poisson' shot noise for the HI-mass-weighted field. (Bottom right) The scale dependence of the bias parameters for the Lagrangian bias model with Zeldovich dynamics. The dashed lines correspond to our default assumption: the scale-independent fits to $k<0.3\hMpc$ data.
    }
    \label{fig:biasperf}
\end{figure}

We begin by evaluating the performance of our bias model using the known initial conditions in the simulation and the aforementioned three metrics, and compare the error power spectrum with the HI-mass-weighted shot noise.  The statistics for the best fit to the HI field at $z=2$ are shown in Fig.~\ref{fig:biasperf}. Though not shown in this figure, the results for $z=4$ and 6 are very similar. Fig.~\ref{fig:biasperf} compares three different bias models: two Lagrangian bias models (as described in \S\ref{sec:biasmodel}), with Zeldovich dynamics and PM dynamics\footnote{Here, our PM dynamics corresponds to a FastPM simulation on a $512^3$ grid with 5 time steps and force resolution $B=1$, i.e.\ a $512^3$ mesh for the force computation. For comparison, the HI data was generated on a $10240^3$ grid with 40 steps to $z=2$ and force resolution $B=2$.} respectively, as well as an Eulerian bias model where the three bias parameters are defined with respect to the fields generated from the Eulerian matter field smoothed at $3\Mpch$ with a Gaussian smoothing kernel.  In addition, we show the simple case of linear Eulerian bias ($b_1^E$) to contrast with our other biasing schemes.

Firstly, comparing the Eulerian and Lagrangian bias models, we find that Lagrangian bias outperforms Eulerian bias at the level of cross-correlation and transfer function on all scales. Lagrangian bias models also lead to quite scale independent error power spectra and much lower noise than the Poisson shot-noise level, while this is not the case for Eulerian models. This implies that we can do analysis with higher fidelity than one would expect from the simplest Poisson prediction and are able to access information to smaller scales. This is in general agreement with the findings of ref.~\cite{Schmittfull19} for halos at different number densities and weightings. Note that while the linear-Eulerian model is a subset of the quadratic-Eulerian model, it performs worse at the level of cross-correlation. This is because the metric for fitting bias parameters is minimizing the noise power spectrum up to $k\simeq 0.3 \hMpc$ where the quadratic Eulerian model is better than the linear bias model.

Amongst Lagrangian bias models we find that the Zeldovich displacements perform better in terms of the cross correlation and the transfer function with our observed HI data field and resolution than the PM dynamics. One can increase the accuracy of the PM displacements by instead increasing the number of time steps or using force resolution $B=2$ for the PM simulations and we find that this improves the model slightly over ZA but makes the modeling much more expensive. Moreover, the difference between the performance of the two dynamics also depends on the resolution of the meshes used to do simulations. Currently, we are modeling the data from a much higher resolution simulation ($20480^3$ force mesh) with a quite `low' resolution ($256^3$ or $512^3$ mesh) PM or ZA dynamics so its not obvious how they should compare, but we do find the performance of PM improving as we increase the resolution of models as might be expected.

Another way to improve the bias model is to add a term corresponding to $b_\nabla^2$ which comes with $k^2$ dependence motivated by peaks bias as well as effective field theory counter-terms. We find that adding such a term does not change the two models at the level of cross-correlation, but significantly improves the PM model over ZA at the level of the transfer function. 
Going one step further, one can also make the bias parameters scale dependent and use them as transfer functions. To assess the scale dependence of the bias parameters, in the lower-right panel of Fig.~\ref{fig:biasperf} we also show the scale dependent bias parameters for the Zeldovich bias model which are fit for by minimizing the error power spectrum independently for every $k$-bin. The best-fit bias parameters still do not have any significant scale dependence up to intermediate scales. As mentioned in previous section, this explains why we find that the fit of bias parameters is quite insensitive to the $k-$range used for fitting the bias model, as long as we do not fit up to highly non-linear scales.

Given the differences in performance of ZA and PM dynamics, how scale-dependent the biases are on the small scales that begin to get increasingly noise dominated in the data and the cost of the ZA vs.~PM forward models, we find the performance of a scale-independent bias model with ZA dynamics to be sufficient for reconstruction. Henceforth, we present results for the reconstruction with this model. However we suspect that it would be worth studying the performance of different bias models in more detail. Our comparison is also likely to change for different number densities and weightings (such as position, mass, HI mass etc.) of the biased tracers.  We leave such a detailed study to future work.

\subsection{Reconstruction}

\begin{figure}
    \centering
    \resizebox{1\columnwidth}{!}{\includegraphics{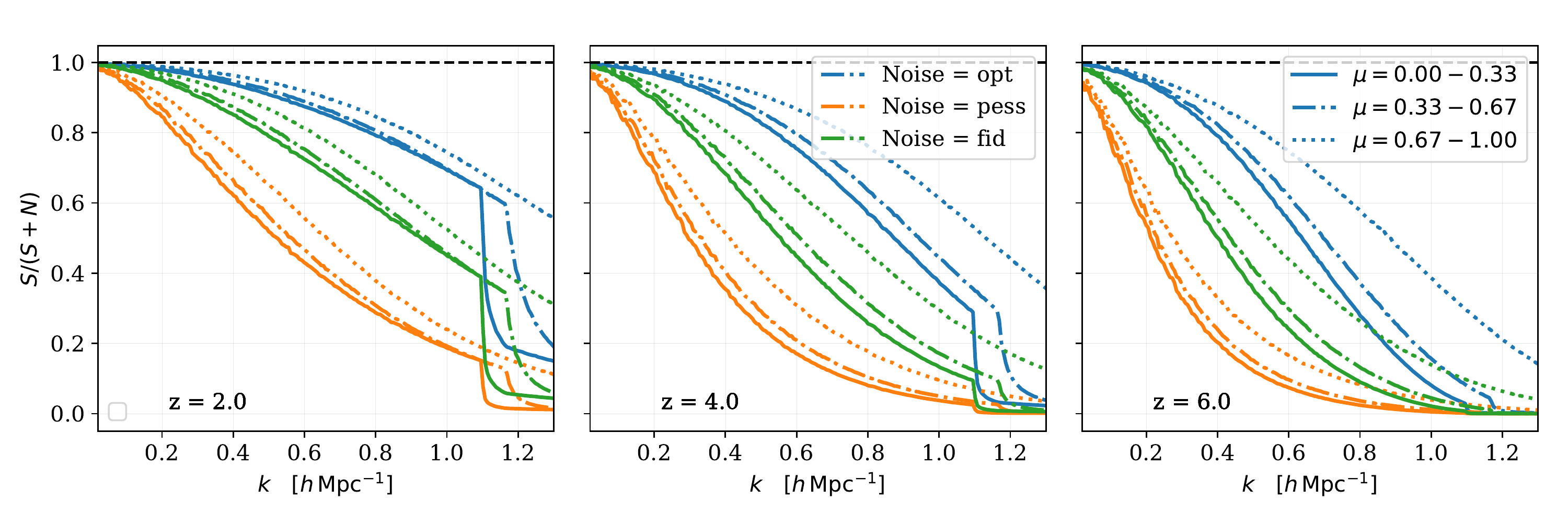}}
    \caption{Ratio of signal to total signal and noise in our HI data at different redshifts as a function of scale and angle ($\mu$ bins) for the three different thermal noise cases considered for reconstruction. Here we have neglected any loss of modes due to foregrounds. 
    }
\label{fig:snr}
\end{figure}

\begin{figure}
    \centering
    \resizebox{1\columnwidth}{!}{\includegraphics{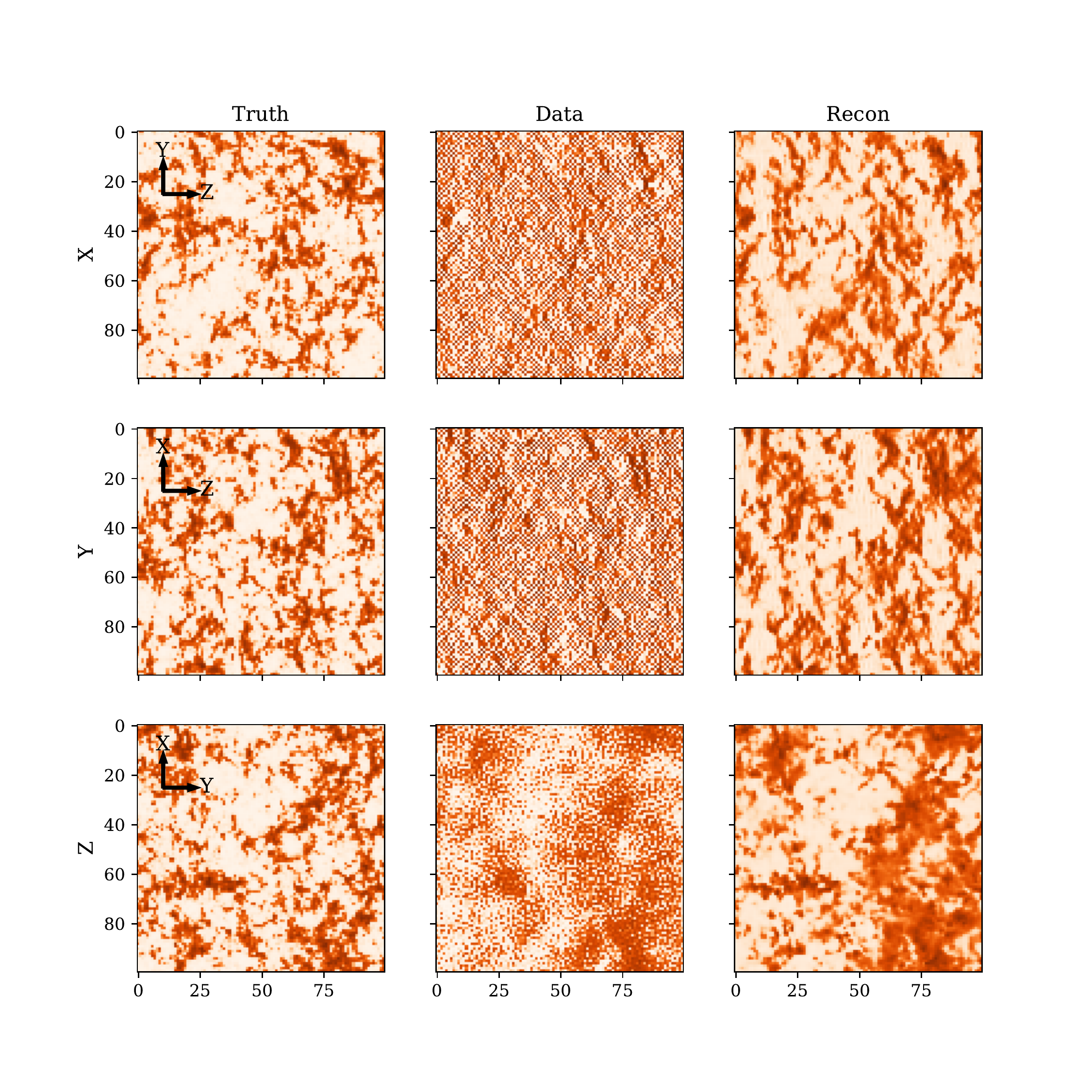}}
    \caption{Different projections for the true HI field, data corrupted with thermal noise and a foreground wedge and our reconstructed HI field at redshift $z=4$ for our fiducial thermal noise corresponding to 5 years of observing with a half-filled array but for a pessimistic wedge ($k_\parallel = 0.03 \hMpc,\, \theta_w=38^\circ$). Different rows show $20 \Mpch$ thick slices when projected along the axis specified by the $y$-axis label. The horizontal and vertical image dimensions correspond to $200 \Mpch$ and correspond to direction specified by arrows on top left for every row. The line of sight of the data is along the $Z$ axis. We use a log-scale color scheme and the color-scale is same for all the panels in the same row. 
    }
\label{fig:image}
\end{figure}

\subsubsection{Configurations}

In this section, we show the results for the reconstruction of the initial and HI field. We do reconstruction on our $1024 \Mpch$ box at redshifts $z=2$, 4 and $6$. The line of sight is always assumed to be along the $z$-axis. Unless otherwise specified, we will show the reconstruction for our fiducial setup which is reconstruction in redshift space, with thermal noise corresponding to $5$ years with a half-filled array, $k_\parallel^{\rm min} = 0.03 \hMpc$ and an optimistic foreground wedge ($\theta_w=5^\circ$, $15^\circ$ and $26^\circ$ at $z=2$, 4 and $6$). To gauge the impact of our assumptions regarding this fiducial setup, we will show comparisons with other setups which include a pessimistic foreground wedge (dashed) corresponding to $\theta_w=15^\circ$, $38^\circ$ and \rsp{$55^\circ$} at $z=2$, 4 and 6 respectively, and different thermal noise configurations corresponding to a full-filled array (optimistic case) and a quarter filled array (pessimistic case).

To gain some intuition about how these noise configurations compare with the signal, we also show the ratio of signal to signal plus noise in Fig.~\ref{fig:snr} as a function of scale and angle for different redshifts. Since the impact of the foreground wedge is binary with respect to scale and angle, we have ignored the loss of modes due to it in plotting Fig.~\ref{fig:snr}.
\rsp{However to emphasize how severe the loss of information due to the wedge is, 
at redshift $z=2$, 4 and 6 we completely loose $21(7)\%$, $60(21)\%$ and $88(40)\%$ of the modes due to the pessimistic (optimistic) foreground wedge.  When we include the thermal noise, even in the fiducial case, a further $30(43)\%$, $24(51)\%$ and $9(37)\%$ of the modes outside the wedge 
become noise dominated.  Thus we preface our results by re-iterating that reconstructing the large-scale modes is a non-trivial problem in these situations.}

\subsubsection{Annealing: Implementation}

To implement our annealing scheme we begin with a $256^3$ mesh and anneal by smoothing the loss-function at smoothing scales corresponding to $4$, 2, 1, 0.5 and $0$ cells. \rsp{For the given resolution, this roughly corresponds to $k\sim 0.2$, 0.4, 0.8 and $1.6\hMpc$. Thus even in the case of largest smoothing, we still have enough small scale modes outside the wedge to inform reconstruction.} We find that the statistics of interest stop changing roughly after $\sim$ 100 iterations, thus we do this many iterations for every step of annealing before declaring it `converged'. After converging on this mesh, we upsample our reconstructed initial field to a $512^3$ grid and repeat our reconstruction exercise starting from this point while comparing to data now painted on this higher resolution grid. At this stage, we do residual smoothing only corresponding to $1$ and 0 cells, 100 iterations each.

We have tried other annealing methods, \rsp{such as different smoothing scales and other upsampling schemes.
In addition, to study convergence, we have also let the optimizer run for longer, increasing the number of iterations.
We find that none of these choices have a significant impact, and while running more iterations can improve results marginally, our aforementioned implementation provides a good balance between computational cost and performance for this work.
Moreover, given the heuristic nature of our annealing scheme, its not obvious if we will converge to a unique solution. To study this, we run simulations with different initial conditions and find that the reconstructed fields are correlated well enough, and hence identical, on the scales of interest here. We establish both of these quantitatively in appendix \ref{app:validanneal}.} 

\subsubsection{Reconstructed field : Visual}

Before gauging our reconstruction quantitatively, we also first visually see the impact of noise and reconstruction at the level of the field to develop some intuition. Fig.~\ref{fig:image} shows different projections of the true HI field, as well as the data which is now corrupted with thermal noise in addition to the foreground wedge and our reconstructed HI field for our fiducial thermal noise corresponding to 5 years of observing with a half-filled array but for a pessimistic wedge ($k_\parallel = 0.03 \hMpc,\, \theta_w=38^\circ$) at $z=4$. The line of sight direction is Z. As expected from Fig. \ref{fig:snr}, where the signal is close to zero on small scales, the data is heavily corrupted with small scale noise and the noise is higher when we take the sum over the transverse direction than when along the line of sight. In fact, visually, one can only faintly distinguish the biggest structure peaks in the corrupted data from noise and its impressive how well we are able to reconstruct smaller structures in the HI field despite this. Due to the foreground wedge, the structures seem to be stretched in the transverse direction in the data. This can also be seen by comparing first two rows with the last, where structures are more isotropic since the stretching is the same in both transverse directions. For the reconstructed field we have visibly reduced this stretching by reconstructing modes in the wedge.  Overall the reconstructed field is smoother than the input data, since we do not fully reconstruct the small-scale modes.

\begin{figure}
    \centering
    \resizebox{1\columnwidth}{!}{\includegraphics{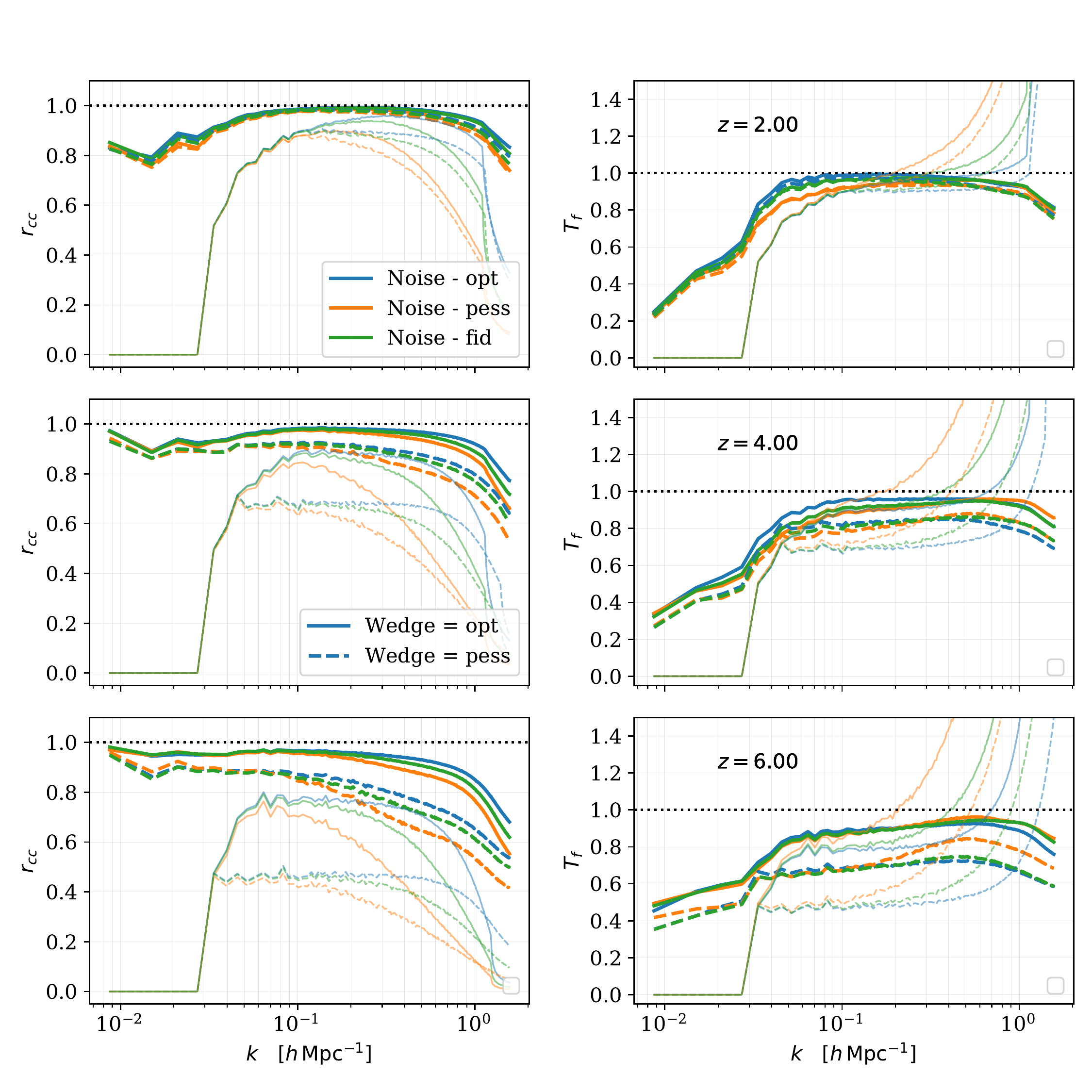}}
    \caption{We show the cross-correlation (left) and transfer function (right) of the reconstructed HI field at $z=2$, 4 and $6$ for three different thermal noise levels as well as two wedge configurations (optimistic and pessimistic; see text for more details). For comparison, we also show these quantities for the noisy and masked data which was used as the input for reconstruction with light colors.
    }
\label{fig:allcompare}
\end{figure}

\subsubsection{Reconstructed field : Two point function}

Next, to quantitatively see how the foreground wedge and thermal noise combined affects our data, we estimate the cross-correlation and transfer function of this noisy input data with the true HI field. This is shown in Fig.~\ref{fig:allcompare} for $z=2$, $4$ and 6 as thin lines. In addition to our fiducial setup, we also consider other configurations for the wedge and thermal noise as outlined at the beginning of this section. This figure contrasts with Fig. \ref{fig:snr} since there we neglected the loss of modes due to foreground wedge and only focused on thermal noise. For the combined noisy data, the transfer function with respect to the true HI field is zero on the largest scales ($k< 0.03\hMpc$) since these modes are completely lost to the foregrounds and we have no information on these scales. On intermediate scales, the cross correlation and transfer function increase but are still well below unity since some modes are still lost to the wedge in every $k-$bin, and as per our expectations, this loss is greater in the pessimistic wedge case. On the smallest scales, the transfer function exceeds one since these modes are dominated by thermal noise. However the cross-correlation on these scales drops rapidly since this noise is uncorrelated to the data and contains no information.

For comparison, we show as thick lines in Fig.~\ref{fig:allcompare} the cross-correlation and transfer function of the reconstructed HI field with true HI field.  This highlights how our reconstruction helps in recovering the information lost due to foregrounds and thermal noise. The gains in cross-correlation coefficient over noisy data are quite impressive, reaching $\sim 0.8$, 0.9, 0.96 for the optimistic wedge on the largest scales for $z=2$, 4 and $6$ respectively where we have access to no modes in the data. The modes in this regime are constructed only out of mode coupling due to no-linear evolution. On intermediate scales, where the data have the most information, the cross correlation reaches $1$ for all redshifts while it drops again (to below $0.8$) on the smallest scales which are thermal noise dominated. A similar trend is observed in the transfer function, where we recover $\sim 40$, $50$ and $60\%$ of the largest scales lost completely to the foregrounds for $z=2$, $4$ and $6$ respectively. We recover more power as we move to smaller scales, with the transfer function reaching close to unity on intermediate scales $k\simeq 0.1\hMpc$ before starting to decrease at the smallest, noise dominated scales.

\begin{figure}
    \centering
    \resizebox{1\columnwidth}{!}{\includegraphics{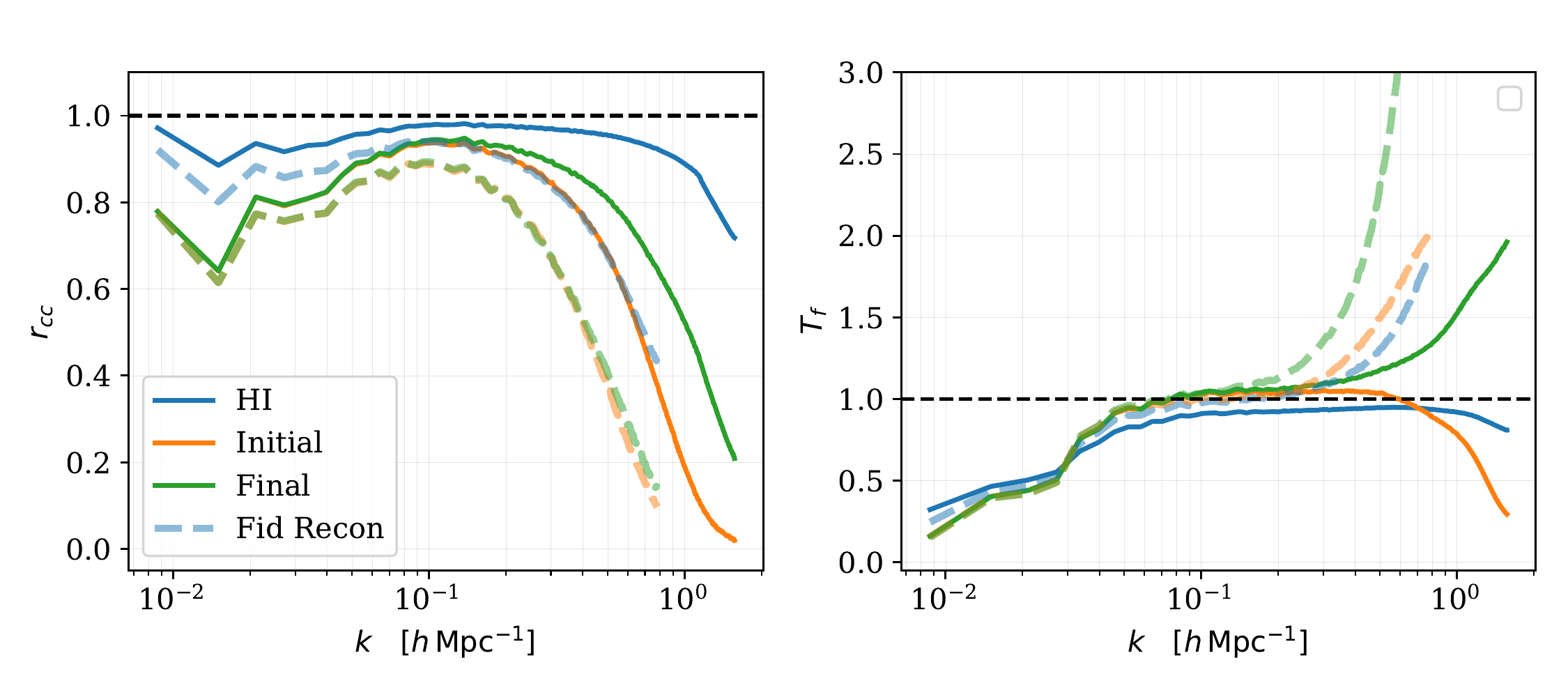}}
    \caption{The cross-correlation (left) and transfer function (right) at $z=4$ for the reconstructed initial matter, final matter and HI field with their corresponding true fields.  We have assumed 5 years of observing with a half-filled array, $k_\parallel^{\rm min} = 0.03 \hMpc$ and $\theta_w=15^\circ$. The dashed lines are for reconstruction on the fiducial $256^3$ mesh while the solid lines are reconstruction after upsampling to a $512^3$ mesh, which leads to significant gains.
    }
\label{fig:allfields}
\end{figure}

Reconstruction is slightly worse in the case of pessimistic wedge, with higher redshifts paying a higher penalty simply due to the larger difference in the two configurations. However the cross-correlation is still $\simeq 80\%$ at $z=6$ on the largest scales, even in the most pessimistic setup. While the foreground wedge affects reconstruction on all scales, thermal noise does not affect reconstruction on the large scales.  On small scales, reconstruction is slightly worse with increasing noise, again penalizing higher redshifts more than lower redshifts.

With our procedure, along with reconstructing the observed data, we also reconstruct the initial and final matter field. Its instructive to see how close are these to the true fields since they have different science applications. The initial (Lagrangian) field can be used to reconstruct Baryon Acoustic Oscillations (BAOs), while the final matter field across redshifts has applications in CMB and weak lensing science. Here we briefly look at the recovery of these fields. In Fig.~\ref{fig:allfields}, we show the cross-correlation and transfer function of the reconstructed initial matter, final matter and HI field for our fiducial setup at $z=4$. As for the HI data field, the cross correlation and transfer function increases as we go from large to intermediate scales since the large scales are completely absent in the data due to foreground wedge. Since the information moves from larger scales to small scales during non-linear evolution, the cross correlation of the reconstructed final-matter and HI field are much better on smaller scales than the initial field. While the transfer function for initial and HI field drops on small scales due to thermal noise, that of final matter field increases over $1$. This is simply because our dynamic model for reconstruction is the Zeldovich approximation while the true data was generated by particle-mesh simulations. 

In Fig.~\ref{fig:allfields} we also show how upsampling improves the performance of reconstruction. The dashed lines show the reconstruction on the fiducial 256$^3$ grid, without any upsampling, while the solid lines show the results when continuing reconstruction after upsampling the reconstructed field to a 512$^3$ mesh.  Since the higher resolution allows us to push to smaller scales it increases the number of modes not dominated by thermal noise while also recovering some of the signal that was earlier lost due to grid-smoothing on these scales.  We find large gains in our reconstruction for all three fields.  While we have not pushed to even higher resolution due to CPU limitations, we suspect it would yield diminishing returns since the smaller scales are dominated by thermal noise. 

\begin{figure}
    \centering
    \resizebox{1\columnwidth}{!}{\includegraphics{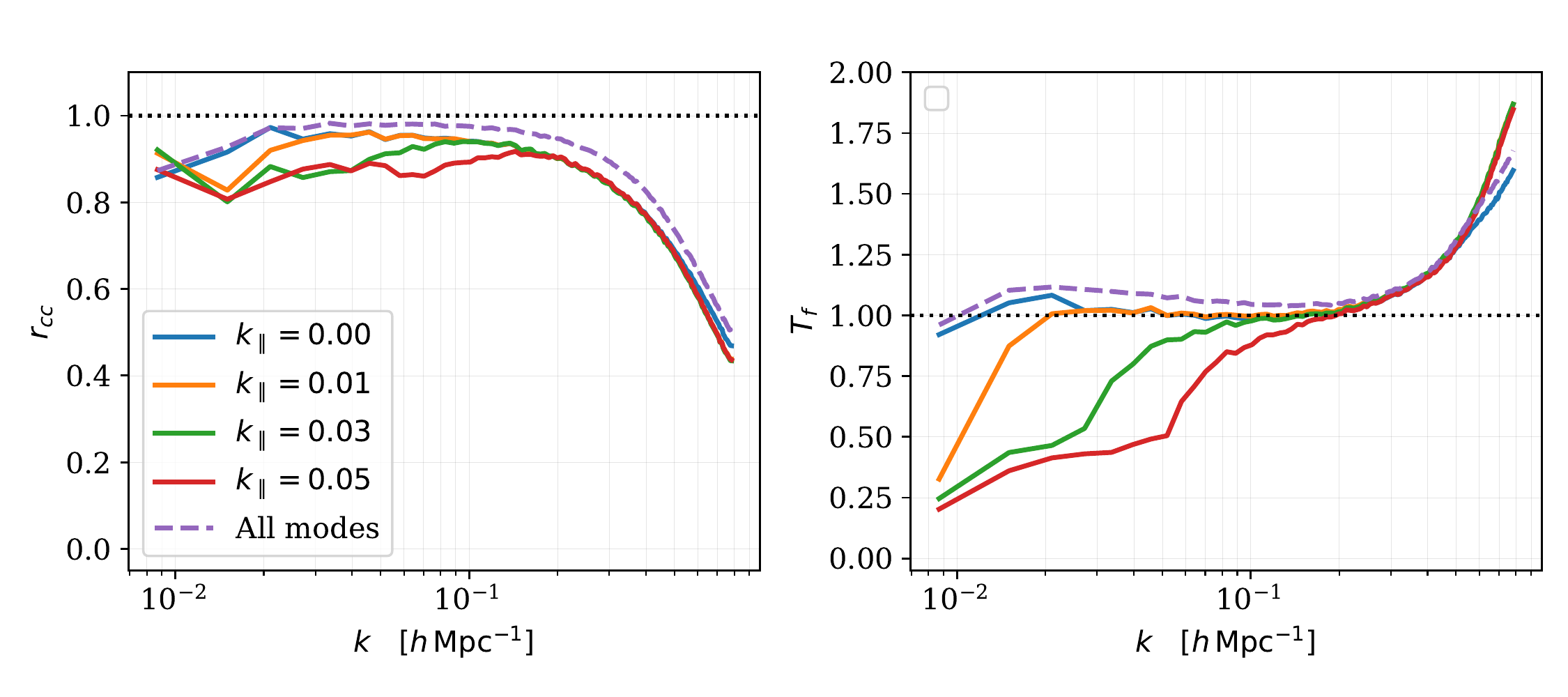}}
    \caption{The cross-correlation (left) and transfer function (right) at $z=4$ with an optimistic wedge ($\theta_w=15^\circ$) and fiducial thermal noise (corresponding to 5 years of observing with a half-filled array) for different $k_\parallel^{\rm min}$ (in $\hMpc$), without upsampling. For comparison, we also show the case where we do not lose any modes to foregrounds (i.e. no wedge and $k_{\parallel}^{\rm min}=0$) but have the same fiducial thermal noise (dashed line).  We have checked that as the thermal noise is reduced the cross correlation moves closer to 1 as expected.
    }
\label{fig:kmin}
\end{figure}
        
\subsubsection{Impact of $k_\parallel^{\rm min}$}
While changing the thermal noise and wedge configurations, we have so far kept the $k_\parallel^{\rm min}$ fixed at $k_\parallel^{\rm min} = 0.03 \hMpc$. In Fig.~\ref{fig:kmin}, we show how the reconstruction performs for different values of $k_\parallel^{\rm min}$, and compare it to the hypothetical case where we loose no modes to foregrounds. Again, we consider our fiducial setup at $z=4$ with optimistic wedge ($\theta_w=15^\circ$) and thermal noise corresponding to 5 years of observing with a half-filled array without upsampling annealing. Reconstruction on all scales is slightly worse than the case when we loose no modes to the foregrounds, but not by much. We find that for a given wedge, small scales are quite insensitive to the $k_\parallel^{\rm min}$ threshold. On large scales, reconstruction gets progressively worse to smaller $k$ as we increase $k_\parallel^{\rm min}$. However even for the most pessimistic case, $k_\parallel^{\rm min} = 0.05\hMpc$, the cross correlation is better than $0.8$ on the largest scales. Furthermore upsampling annealing would improve the reconstruction on all scales.

\subsubsection{Real vs.~redshift space}

\begin{figure}
    \centering
    \resizebox{1\columnwidth}{!}{\includegraphics{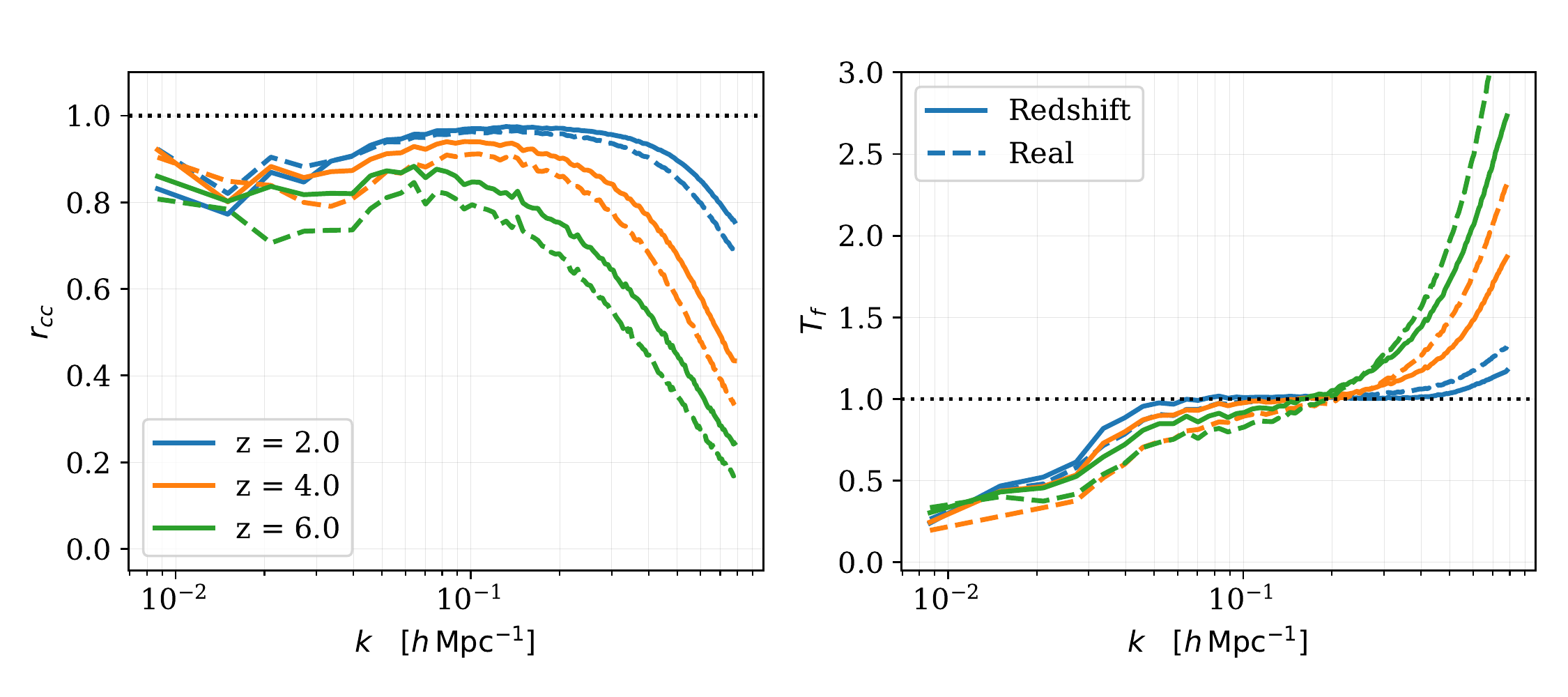}}
    \caption{We show the cross-correlation (left) and transfer function (right) of the reconstructed HI field at $z=2$, 4 and $6$ for the fiducial noise setup and optimisitic wedge in real and redshift space.  The additional information available in the redshift-space field enhances the recovery of the signal.
    }
\label{fig:realrsd}
\end{figure}

Its also instructive to compare reconstruction in real and redshift space, since this gives us access to new signal (through the velocity field) but it is often the case that one loses power in Fourier modes in the radial direction. However for the 21-cm signal finger-of-god effects are subdominant and most of the redshift space signal can be modeled with perturbation theory \cite{VN18,HiddenValley19}.  As a result, we find that doing reconstruction from the redshift space data improves our results over real space data.  We show this in Fig.~\ref{fig:realrsd} where we compare the statistics for reconstruction in real space (dashed) and redshift space (solid) for $z=2$, $4$ and $6$. We model the redshift space data by moving the Lagrangian fields to Eulerian space with Zeldovich dynamics, as before, and then using the Zeldovich velocity component along the line of sight. The velocity field information residing in the anisotropic clustering provides additional information which improves our performance in redshift space over real space. The gains are largest at $z=6$ and decrease with decreasing redshift. This is likely because we model only the linear dynamics while non-linear RSD, as well as the finger-of-god effects, increase at lower redshifts.  Using higher order perturbation theory, or the particle mesh dynamics, should improve the performance at lower $z$.

\subsubsection{Angular cross-correlation}

\begin{figure}
    \centering
    \resizebox{1\columnwidth}{!}{\includegraphics{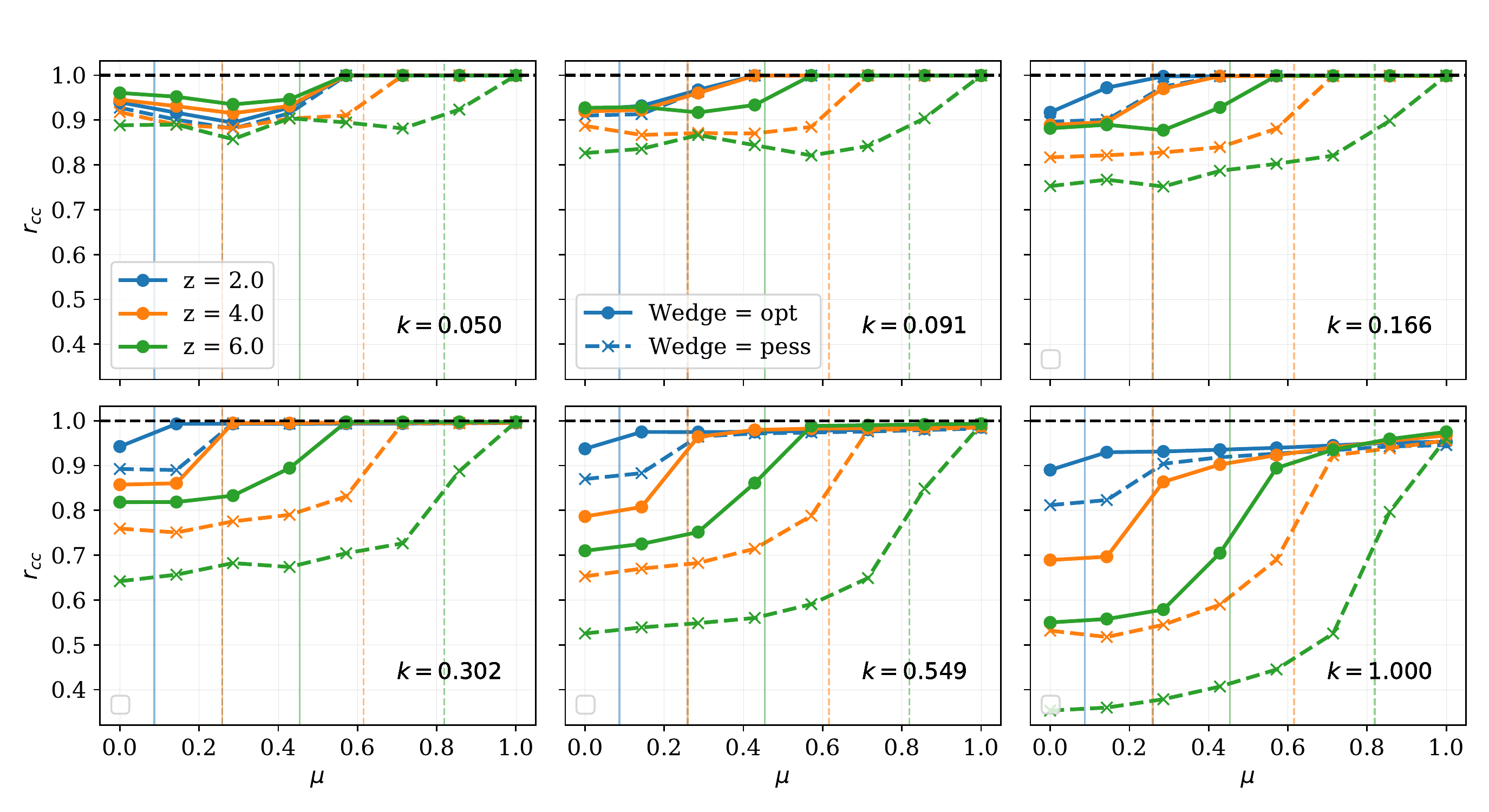}}
    \caption{We show the cross-correlation of the reconstructed HI field with the corresponding true field as a function of the angle $\mu$ with the line of sight in different $k$ bins for fiducial thermal noise and two wedge configurations (with corresponding $\mu_{\rm wedge}$ shown as thin vertical lines) at $z=2$, 4 and $6$. All the modes are reconstructed well along the line of sight, while the large scales are reconstructed better than small scales in the transverse direction. 
    }
\label{fig:compare2dmu}
\end{figure}

Given that the line of sight and transverse modes are differently affected by foreground wedge, thermal noise and redshift space distortions, we conclude this section by looking at the cross-correlation coefficient for the reconstructed data as a function of $\mu$. This is shown in Fig.~\ref{fig:compare2dmu} for our fiducial thermal noise model (after annealing).  We recover the largest scales almost perfectly, even though we do not have any modes along the line of sight in the data.  On the other hand small scale modes along the line of sight are significantly better reconstructed than the transverse modes.  We find that the reconstruction of line-of-sight modes is less sensitive to the loss of modes in the foreground wedge than transverse modes. However in all cases the better foregrounds are controlled, and the smaller the wedge can be made, the better reconstruction proceeds.


\section{Implications}
\label{sec:implications}

Our ability to reconstruct long wavelength modes of the HI field has several implications, and we explore some of them here.
The first two subsections discuss cross-correlation opportunities afforded by reconstruction while the last describes the improvement in BAO distance measures.

The utility of large 21-cm intensity mapping arrays is the highest in the high redshift regime, where there is a lot of cosmic volume to be explored and which is the most difficult to be accessed using other methods. While there are some fields that will cross-correlate straight-forwardly with the 21-cm data, notably the Lyman-$\alpha$ forest and sparse galaxy and quasars samples which entail true three-dimensional correlations, any field that is projected in radial direction occupies the region of the Fourier space that is the most contaminated with the foregrounds. This technique allows us to re-enable these cross-correlations and thus significantly broadens the appeal of high-redshift 21-cm experimentation. We discuss two important examples below.

The first question that needs to addressed is whether it is preferable to cross-correlate with the reconstructed HI field or the evolved matter field or even the linear field. While there are some arguments against, we opted to use the reconstructed HI field, which we refer to simply as the ``reconstructed'' field. Most importantly, on scales where data are available, this field will resemble the measured data regardless of modeling imperfections.  In particular, Figure \ref{fig:kmin} show that the HI field has a higher cross-correlation coefficient than the initial field, indicating that the modeling is imperfect but nevertheless sufficiently flexible to account for these deficiencies.

Finally we note that low-$k$ modes are extremely important in the quest for Primordial Non Gaussianities (PNG), with constraints on PNG from 21\,cm survey severely hampered by the loss of long wavelength modes due to foregrounds \cite{CVDE-21cm}. PNG of the local-type would  benefit enormously from reconstruction, which appears to be most robust way to recover the true signal at large scales free of residual foregrounds \cite{2004PhR...402..103B,2014PTEP.2014fB105T}. The equilateral and squeezed triangle configurations could potentially also benefit, because their sensitivity is normally limited by non-linear gravitational evolution that produces similarly shaped bispectra. We intend to return to this specific case in future work.

\subsection{Redshift distribution reconstruction}

A common problem facing future photometric surveys is to determine the redshift distribution of the objects, many of which may be too faint or too numerous to obtain redshifts of directly \cite{Newman15}.  One approach is to use `clustering redshifts', wherein the photometric sample is cross-correlated with a spectroscopic sample in order to determine $dN/dz$ of the former \cite{Ho08,Newman08,Erben09,Menard13,McQuinn13}.

One difficulty with this approach for intensity mapping surveys is that, for broad $dN/dz$, the photometric sample only probes $k_\parallel\approx 0$.  In linear theory the cross-correlation between a high-pass filtered 21-cm field and the photometric sample is highly suppressed.
Translating a redshift uncertainty of $\delta z$ into a comoving distance uncertainty of $\sigma_\chi=c\,\delta z/H(z)$, to probe $k_\parallel=0.03\hMpc$ requires $\delta z/(1+z)<0.01-0.015$ at $z=2-6$.  Such photo-$z$ precision is in principle achievable, given enough filters, but primarily at lower redshift and for brighter galaxies \cite{Gorecki14,Laigle16}.  The more common assumption of $\delta z/(1+z)=0.05$ corresponds to
\begin{equation}
    \sigma_\chi \simeq 120\,h^{-1}{\rm Mpc} \left(\frac{1+z}{5}\right)^{-1/2}
    \label{eq:scatter}
\end{equation}
for $2\le z\le 6$.  Modes with $k_\parallel=0.03\hMpc$ are almost entirely unconstrained by such measurements.

To be more quantitative we note that such a photometric redshift uncertainty would smooth the galaxy field in the redshift direction.  At very low $k$, and for the purposes of exploration, we can assume scale-independent linear bias so that on large scales we have
\begin{equation}
    \delta_{\rm photo}(k,\mu) =  \left( b_p + f\mu^2\right)  \mathcal{D}
                                  \, \delta_m(\vec{k}) + {\rm noise} \quad , \quad  \mathcal{D}(k,\mu)\equiv \exp[-k^2\mu^2\sigma_\chi^2/2]
    \label{eq:photo}
\end{equation}
where $\delta_m(\vec{k})$ is the matter overdensity and $\mathcal{D}$ encompasses the effect of photometric redshift uncertainty which we have approximated as Gaussian. Similarly, for the true HI field, assuming scale independent bias on these scales, 
\begin{equation}
    \delta_{\rm HI}(k,\mu) =  \left( b_{\rm HI} + f\mu^2\right)  \, \delta_m(\vec{k}) + {\rm noise}
\end{equation}

Using Eq.~\ref{eq:rt-def} the cross-power spectrum with $\delta_{\rm rec}$ is then given by
\begin{equation}
    P_\times(k,\mu)\equiv
    \left\langle\delta_\rec \delta_{\rm photo}\right\rangle
    = \frac{b_p + f\mu^2}{b_\HI + f\mu^2}
    \,\mathcal{D}(k,\mu)\ r(k,\mu)T_f(k,\mu)\ P_{HI}(k,\mu)
\label{eq:cross-with-recon}
\end{equation}

At high redshift both the galaxies and HI are highly biased and we are interested in $\mu\approx 0$ modes so the first factor becomes $b_p/b_{HI}$ which simply rescales the amplitude of $T_f$. Eq.~\ref{eq:cross-with-recon} shows why we need to reconstruct the low $k$ modes in order to achieve a significant signal in cross-correlation: $\mathcal{D}$ highly damps the signal at high $k_\parallel$ and without reconstruction $rT_f\to 0$ at low $k_\parallel$.

For Gaussian fields the variance of the cross-correlation is 
${\rm Var}\left[\delta P_\times\right] = N_{\rm modes}^{-1} \left( P_\times^2 +  P_{\rm photo}P_{\rm rec}\right)$
where $N_{\rm modes}$ is the number of independent modes in the bin and $P_{\rm rec} = T_f^2 P_{HI}$ is the reconstructed auto-spectra of HI field along with the reconstruction noise. Additionally, the auto-spectrum $P_{\rm photo}$ includes a contribution from the noise auto-spectrum, which we assume to be shot-noise: $\bar{n}^{-1}$.  In this limit
\begin{equation}
    P_{\rm photo} = \left( b_p + f\mu^2\right)^2  \mathcal{D}^2 P_m + \frac{1}{\bar{n}}
\end{equation}
where $P_m$ is the matter power spectra. Thus,
\begin{equation}
    \frac{{\rm Var}[P_\times]}{P_\times^2} \simeq N_{\rm modes}^{-1}
    \left( 1 + \rho^{-2} \right)
\end{equation}
where  
\begin{equation}
    \rho^2 = \frac{P^2_\times}{P_{\rm photo}P_{\rm rec}} =  r^2 \frac{b_p^2\mathcal{D}^2\,\bar{n}P_m}{1+b_p^2\mathcal{D}^2\,\bar{n}P_m}
    \label{eq:rho}
\end{equation}
quantifies the effective signal and plays the role of the more familiar $\bar{n}P$ that frequently appears in power spectrum errors \cite{FKP94}. An important caveat here is that to estimate $\rho$ as a function of $\mu$, the correct way is to integrate over the $\mu$-bin, and not simply to estimate it at the bin-center. Since the photometric damping kernel scales as $\mu^2$, estimating at the bin center causes over damping and the impact is especially severe for the smallest $\mu$ bin as modes with $\mu \rightarrow 0$ should effectively be undamped.

\begin{figure}
    \centering
    \resizebox{1\columnwidth}{!}{\includegraphics{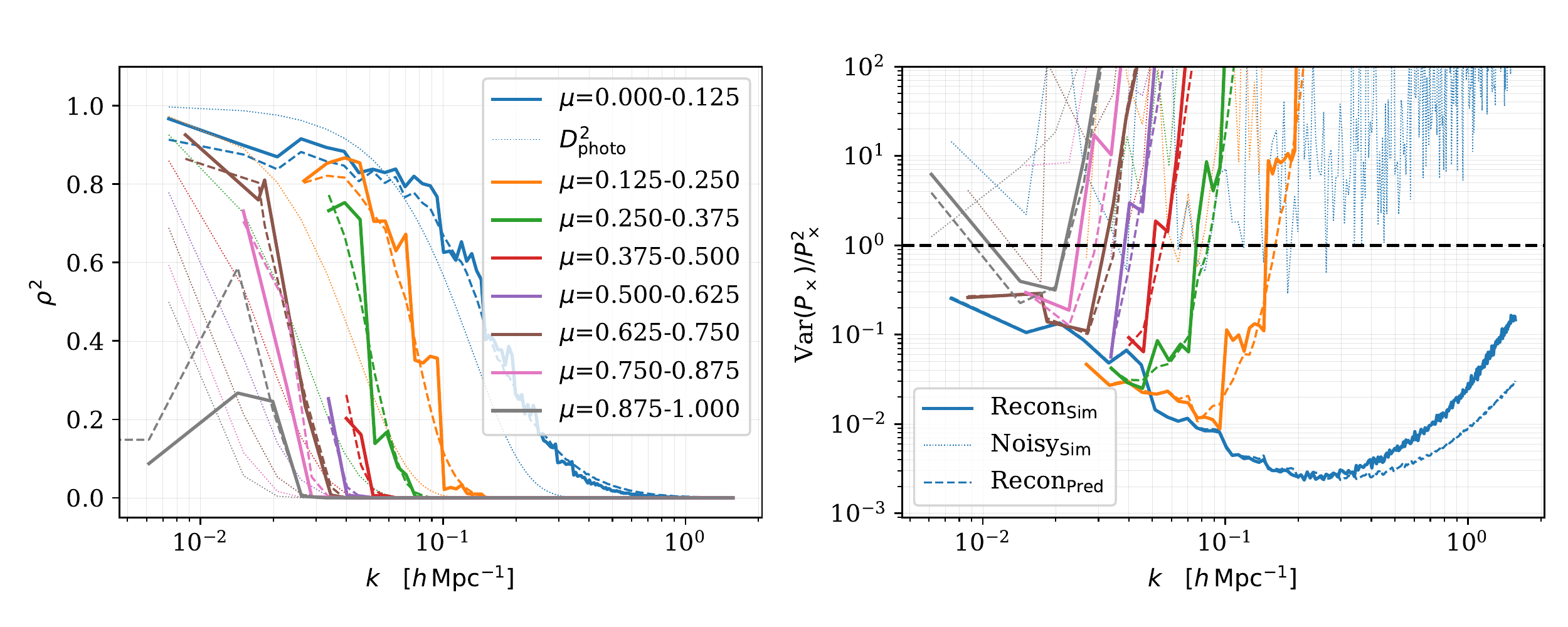}}
    \caption{We show the effective cross-correlation ($\rho^2$, left) as well as the variance of the cross-spectra (right) of the reconstructed HI field (fiducial setup) with the photometric field at $z=4$ and number density $\bar{n}\simeq 10^{-2.5}(\hMpc)^3$, for different $\mu$-bins. The solid lines show the measurement from the simulations while the dashed lines are theoretical predicition based on Eq.~\ref{eq:rho}. The dotted lines on the left show the photometric damping kernel of Eq.~\ref{eq:photo} while on the right they show the variance for the input HI data with thermal noise and wedge (see text for discussion).
    }
\label{fig:photo}
\end{figure}

For clustering redshifts with dense spectroscopic samples most of the weight comes from transverse modes near the non-linear scale.  For example, the optimal quadratic estimator for $dN/dz$ (\cite{McQuinn13}; \S\,5.1) weights $P_\times(k_\perp,k_\parallel\approx 0)$ with $P_{\rm photo}^{-1}(k_\perp,0)$ which typically peaks at several Mpc scales.  Similarly the estimator of ref.~\cite{Menard13} has weights proportional to $J_0(kr)$ integrated between $r_{\rm min}$ and $r_{\rm max}$.  Taking $r_{\rm min}$ to be large enough that 1-halo terms are small leads to similar scales being important.  As shown in Fig.~\ref{fig:compare2dmu} the transverse modes at intermediate scales are quite well reconstructed by our procedure, suggesting that 21-cm experiments would provide highly constraining clustering redshifts even for very high $z$ populations.

As an example to show how well we are able to cross-correlate photometric surveys with the reconstructed HI field, we consider a strawman survey at redshift $z \approx 4$, as outlined in ref.~\cite{Wilson19}, that detects Lyman break galaxies (LBG) with the $g$-band dropout technique. Based on Table 5 of ref.~\cite{Wilson19} we take $\bar{n}\simeq 10^{-2.5}(\hMpc)^3$ and $b\simeq 3.2$.  Given these numbers, we show how the $\rho^2$ as well as noise-to-signal ratiofrom  for the cross-spectra as function of scales and angles in Fig.~\ref{fig:photo}. The solid lines are the estimates from the simulations while the dashed lines are the `theoretical' predictions. To generate the photometric data in the simulation, we simply select the heaviest halos up to the given number density. To implement photometric uncertainties, we  smooth the data with the Gaussian kernel of Eq.~\ref{eq:photo}. An alternate way to implement photometric redshifts would be to scatter the positions of halos along the line of sight with standard deviation given by Eq.~\ref{eq:scatter}, and we find that this leads to similar results for scales where ${\rm Var}[P_\times]/P_\times^2 < 1$, but becomes noisier on smaller scales. Thus here, we stick with the smoothing implementation. Given a photometric field, we then estimate its auto-spectra (with shot-noise ) and cross-spectra with the reconstructed HI field in $\mu$-bins and use Eq.~\ref{eq:rho} to estimate $\rho$ and correspondingly the variance. Similarly, to get the `theoretical' prediction, we linearly interpolate the estimated $r_c$ for every $k$-bin as function of $\mu$ and then integrate the last term of Eq.~\ref{eq:rho} in every $\mu$-bin. 

For the reconstructed HI field, we find in Fig.~\ref{fig:photo} that the signal to noise in both, the predicted and measured cross-spectra for the smallest $\mu$-bin is of the order 10 on all scales while reaching $\sim 100$ on the intermediate scales that are reconstructed the best. For higher $\mu$-bins, the signal to noise is still of the order $\sim 10$ on the largest scales but it deteriorates rapidly due to the photometric damping kernel. For comparison, we also the show as dotted lines the signal to noise for the cross-spectra with the input noisy HI data (with the wedge and the thermal noise) and see that it does not achieve 1 on any scales. This clearly demonstrates the gains made by using reconstructed HI field for estimating photometric redshifts by measuring clustering. 

\subsection{Cross-correlation with CMB weak lensing}

In CMB lensing we are attempting to cross-correlate our HI with the reconstructed convergence field, $\kappa$, given by
\begin{equation}
  \kappa(\theta) = \int_0^{\chi_s} d\chi\ W_\kappa (\chi) \delta_m(\chi,\theta)
  \quad{\rm with}\quad
  W_\kappa = \frac{3 \Omega_m (1+z)}{2} \left(\frac{H_0}{c}\right)^2 \frac{\chi(z) (\chi_s-\chi)}{\chi_s}
  \quad .
\end{equation}
with $\chi_s\simeq \chi (z=1150)$.  Lensing kernel varies very slowly in radial direction and hence it will be insensitive to parallel wavenumbers larger than $\kpar\sim 10^{-3} h\,{\rm Mpc}^{-1}$. Therefore, the suppression is even more strong than in the case of photometric redshifts and cross-correlation is completely hopeless unless one recovers the $\kpar\approx 0$ modes.

When cross-correlating with a tomograpic bin of reconstructed 21-cm density, the angular cross power spectrum is given by
\begin{equation}
C_\ell^\times = \int_{z_{\rm min}}^{z_{\rm max}} \frac{c\, dz}{\chi^2(z) H(z)} W_\kappa(z)
\ r(z,\kperp(z),\kpar=0) T_f(z,\kperp(z),\kpar=0) b_\HI (z) P_{mm}(\kperp(z),z)
\end{equation}
where transverse wavenumber is evaluated at $\kperp=\ell/\chi(z)$ and we have neglected any magnification bias.  For slices that are thin $\Delta \chi \ll \chi$ (but thick enough so that the Limber approximation is valid) this simplifies to
\begin{equation}
  C_\ell^\times = W_\kappa\ b\ r T_f \frac{c\,\Delta z}{\chi^2 H(z)}  P_{mm} 
\end{equation}
Similarly, the HI auto power is given by
\begin{equation}
  C_\ell^{HI} = b^2 T_f^2 \frac{c\,\Delta z}{\chi^2 H(z)}  P_{mm} 
\end{equation}
while the $\kappa$ autospectrum remains an integral over the line of sight.
As before, the noise ${\rm Var}[C_\ell^\times] = N_{\rm modes}^{-1}((C_\ell^\times)^2 + C_\ell^{\kappa\kappa} C_\ell^{\HI}) \propto r^2$ assuming the second term dominates.  The amount of signal-to-noise available for extraction is thus proportional to $r$.  Our method thus allows us to reconstruct over 80\% of all signal available (at a fixed noise of CMB map) compared to none in case of using the pure foreground cleaned 21-cm signal.

\subsection{BAO reconstruction}
\label{sec:BAOrecon}

A major goal of high $z$, 21-cm cosmology is the measurement of distances using the baryon acoustic oscillation (BAO) method \cite{Weinberg13}.  In ref.~\cite{HiddenValley19} we found that future interferometers would be able to make high-precision measurements of the BAO scale out to $z\simeq 6$ with scale-dependent biasing and redshift-space distortions having only a small effect on the signal.  Non-linear structure formation damps the BAO peaks and reduces the signal-to-noise ratio for measuring the acoustic scale \cite{Bharadwaj98,Meiksin99,ESW07,Smith08,Crocce08,Matsubara08a,Seo08,White14,White15}. Since the non-linear scale shifts to smaller scales at high-redshifts, this damping is modest. We find that only the fourth BAO peak is slightly damped (compared to linear theory) at $z=2$ and no damping is visible at $z=6$.

Galaxy surveys which measure BAO typically apply a process known as reconstruction \cite{ES3, Obuljen17} to their data in order to restore some of the signal lost to non-linear evolution (e.g.~see refs.~\cite{BOSS_DR12,Carter18,Bautista18} and references therein).  It is known that the absence of low $k_\parallel$ modes and the presence of the foreground wedge make reconstruction much less effective \cite{SeoHir16}, though some of the lost signal can be recovered with other surveys \cite{Cohn16}.  Our approach provides another route to BAO reconstruction \cite{Modi18}, so it is interesting to ask how the performance of the algorithm compares to standard reconstruction and how it is impacted by loss of data due to foregrounds. For standard reconstruction, we will follow the algorithm outlined in ref.~\cite{SeoHir16}. This differs slightly from the traditional reconstruction using galaxies in that one removes the modes lost in the wedge to estimate Zeldovich displacements and instead of point objects like galaxies, one shifts the HI fields using this displacement. Since BAO reconstruction is most effective at $z=2$, we focus our attention on this case -- which is also the epoch at which our knowledge of the manner in which HI inhabits halos is most secure.  At this redshift an instrument like PUMA is shot-noise limited.

\begin{figure}
    \centering
    \resizebox{1\columnwidth}{!}{\includegraphics{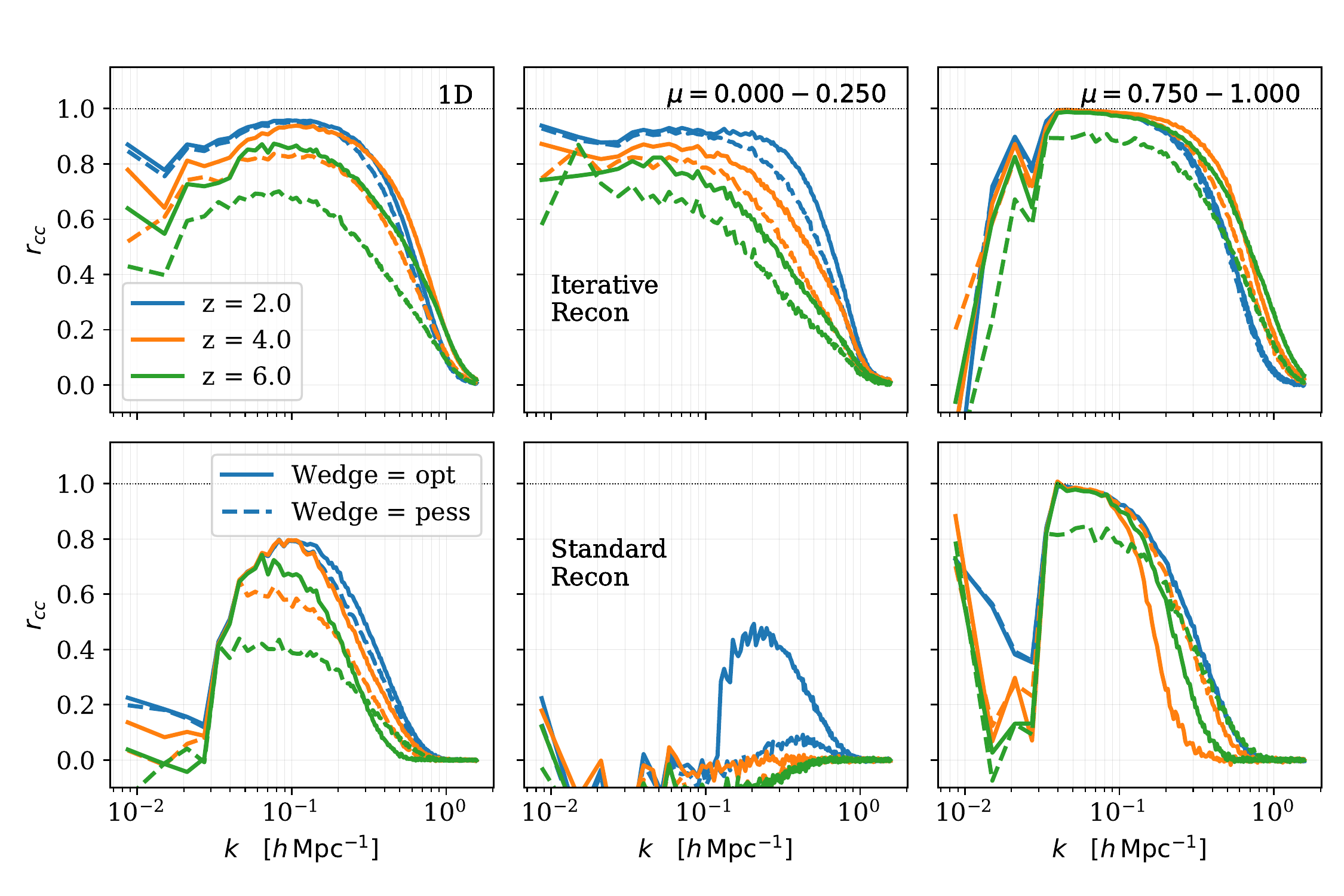}}
    \caption{We show the cross-correlation of the reconstructed initial/linear field with the true Lagrangian field for our method of reconstruction (iterative, top row) and standard reconstruction (bottom row) for the three redshifts and two wedge configurations (solid and dashed) and fiducial thermal noise setup corresponding to 5 years with half-filled array. The first column is the  cross correlation monopole while the second and third column show the cross correlation in bins nearly perpendicular to and along the line of sight respectively.
    }
\label{fig:rccstd}
\end{figure}

To gauge how well are we able to recover the BAO signal we closely follow ref.~\cite{Modi18} which in turn builds upon refs.~\cite{Seo08, Seo16}.  Specifically we employ a simple Fisher analysis to estimate the uncertainty in locating the BAO features which allow us to measure the sound horizon at the drag epoch, $s_0$.  This parameter is only sensitive to the BAO component of the power spectrum, which is damped due to Silk damping and non-linear evolution. The information lost in the latter is recovered with reconstruction and we quantify its success by measuring the linear information in the reconstructed field, $\delta_r$. This is done by estimating its projection onto the linear field, $\delta_{\rm lin}$ in the form of the `propagator'
\begin{equation}
    G(k, \mu) = \frac{\langle \delta_r \delta_{\rm lin} \rangle}{b\langle \delta_{\rm lin} \delta_{\rm lin} \rangle} = \frac{r_c T_f }{b}
\end{equation}
where $b$ is the bias of the field and $r_c$ and $T_f$ are the corresponding cross correlation and transfer function of the reconstructed field with the linear field. Thus, the linear `signal' in the field is $S = b^2 G^2 P_{\rm lin}$ while under the Gaussian assumption of fields, the total variance is given by the square of the power spectra of the field itself. As a result, the total signal to noise for the linear information is:
\begin{equation}
    \frac{S}{N} = \frac{b^2 G^2 P_{\rm lin} }{\langle \delta_{r} \delta_{r} \rangle} = r_c^2
\end{equation}

We compare the performance of the two methods in Fig. \ref{fig:rccstd} where we show the cross-correlation of the two reconstructed fields with the true linear field for different redshifts and wedge configurations. Since the comparison for other noise levels is qualitatively similar, we show only the case of fiducial setup corresponding to 5 years of observation with half-filled array. Overall, our method outperforms standard reconstruction significantly, with higher cross-correlation on all scales and extending to smaller scales. More importantly, as shown in the second column of Fig.~\ref{fig:rccstd}, standard reconstruction fails to reconstruct modes inside the foreground region while our method is able to do so. These modes provide complementary information to the line of sight modes, constraining the angular diameter distance while line of sight modes constrain $H(z)$. Thus, unlike the standard method, our reconstruction should be able to constrain the angular diameter distance as well as improve measures of the Hubble parameter.

To be more quantitative, we can use the Fisher formalism to estimate the error on $s_0$ as (for a derivation, see refs.~\cite{Seo08,Seo16}) 
\begin{equation}
F_{{\rm ln}\, s_0} = \bigg(\frac{s_0}{\sigma_{s_0}}\bigg)^2 =  V_{\rm{survey}} A_0^2 \int_{k_{\rm min}}^{k_{\rm max}}k^2 dk \ \frac{b^4 G^4(k)\,\exp[-2(k \Sigma_s)^{1.4}]}{\left[P_{\rm lin}(k)/P_{0.2}\right]^2}  
\label{eq:Fishers0}
\end{equation}
where we have integrated over the angles assuming isotropy, $V_{\rm survey}$ is the survey volume, $A_0$ = 0.4529 for WMAP1 cosmology, $\Sigma_s\simeq 7.76\,h^{-1}$Mpc is the Silk damping scale and $P_{0.2}$ is the linear power spectrum at $k=0.2\,h\,{\rm Mpc}^{-1}$. While the nature of this calculation is rather crude, our aim here is to not do any accurate Fisher forecast for measuring BAO, but simply to broadly compare our reconstruction with the standard reconstruction under a sensible metric and to that end this procedure should suffice. To be more yet conservative, instead of quoting the relative Fisher errors individually, we estimate the ratio of predicted errors for the two methods. We find that for the BAO peak, $s_0$, under the fiducial thermal noise setup, our reconstruction reduces errors over the standard method by a factor of $\sim 2$ ($2$) and $\sim 2.8$ ($3.5$) for optimistic (pessimistic) wedge at $z=2$ and $6$. The conclusions remain relatively unchanged for other thermal noise configurations. While this is for angular averaged BAO, as mentioned previously, we expect gains to be comparable for Hubble parameter and larger for angular diameter distance. We leave a more accurate calculation for future work.

\section{Conclusions}
\label{sec:conclusions}

In this paper, we have applied recently developed field reconstruction methods to the case of future 21-cm cosmology experiments. In many respects this is an optimal target for such methods. First, we are dealing with continuous fields which are naturally more suited for these methods, since the problem of the non-analytical object creation does not apply. \rsp{
In galaxy clustering, while its possible to circumvent the discreteness of the data by averaging in pixels that are sufficiently big that the galaxy counts in cells become effectively continuous, it leads to information loss. However, for 21-cm intensity mapping experiment, the finite resolution of the experiment naturally leads to a continuous version of the problem. We simply do not have information of scales that are small enough to even contemplate separating individual objects, even though the signal is dominated by them.}

Second, the 21-cm intensity field is dominated by numerous contributions from relatively low mass dark matter halos \cite{Castorina17,VN18,HiddenValley19}, which are well described by a low-order bias expansion to relatively high wavenumbers. \rsp{This allows us to start from the underlying dark matter field and model the observed data at the field level more simply and accurately than for discrete and more biased tracers such as galaxies.} 
Third, we are missing some regions of $k$-space (\rsp{$\sim 20(7)-80(40)\%$ of the modes for our pessimistic (optimistic) case}) but, on the other hand, measuring a very large number of modes over other regions of $k$-space. As we have demonstrated, the measured modes \rsp{and the couplings introduced by non-linear evolution} more than make up for the missing ones.

Our method proceeds by maximizing \rsp{the posterior for the initial conditions, which is constructed by combining a Gaussian prior on the initial field with the likelihood of a forward model matching the observations}. In our case the observations are the mock, redshift space $\delta_{HI}$ in $k$-space, as would be measured by a 21-cm interferometric survey.  The forward model consists of a quadratic, Lagrangian bias scheme paired with non-linear dynamics which can be either perturbative dynamics or a particle mesh simulation. We find that simple Zeldovich dynamics for such a model do a good job of fitting our mock data, with errors well below the shot noise level in the field (Fig.~\ref{fig:biasperf}).

For $2<z<6$ we are able to recover modes down to $k\simeq 10^{-2}\hMpc$ with cross-correlation coefficients larger than 0.8 for both optimistic and pessimistic assumptions about foreground contamination (Fig.~\ref{fig:allcompare}).  Our reconstruction is relatively insensitive to loss of line-of-sight modes, up to $k_\parallel^{\rm min}=0.05\hMpc$, but more sensitive to missing modes in the `wedge' (\S\ref{sec:instruments}) as shown in Fig.~\ref{fig:kmin}.  For our fiducial thermal noise assumptions we recover the $k\simeq 10^{-2}\hMpc$ modes with cross-correlation coeffecient greater than $0.9$ in all directions and the line of sight modes almost perfectly, even though we do not have any of these modes in the data. On the other hand small scale modes along the line of sight are significantly better reconstructed than the transverse modes. Thus as shown in Fig. \ref{fig:compare2dmu}, we find that the reconstruction of line-of-sight modes is less sensitive to the loss of modes in the foreground wedge than transverse modes. However in all cases the better foregrounds are controlled, and the smaller the wedge can be made, the better reconstruction proceeds.  At $z\simeq 2$ our reconstructions are relatively insensitive to thermal noise over the range we have tested.  However at higher $z$ increasingly noisy data leads to steadily lower cross-correlation coefficient, as shown in Fig.~\ref{fig:allcompare}.

Our method also provides a technique for density field reconstruction for baryon acoustic oscillations (BAO) \cite{Modi18}. It is known that the absence of low $k_\parallel$ modes and the presence of the foreground wedge make reconstruction much less effective \cite{SeoHir16}, though some of the lost signal can be recovered with other surveys \cite{Cohn16}.  By constraining the reconstructed initial field, as well as the evolved HI field, we can use the measured modes to `undo' some of the non-linear evolution and increase the signal-to-noise ratio on the acoustic peaks (see \S\ref{sec:BAOrecon}). In particular, we find that while the standard methods of reconstruction recover almost no modes in the foreground wedge, our method reconstructs modes with cross correlation better than $80\%$. This can lead to improvements of a factor of 2-3 in isotropic BAO analysis and more in transverse directions. We note that \rsp{these gains over standard methods are significantly larger than in the case of galaxy surveys since the standard reconstructions is already quite efficient for the latter \cite{SeoHir16, Schmittfull17, Modi18}}.

We have focused primarily on the reconstruction technique in this paper, but the ability to reconstruct low $k$ modes opens up many scientific opportunities.  We described (\S\ref{sec:implications}) how 21-cm fluctuations could provide superb clustering redshift estimation at high $z$, where it is otherwise extremely difficult to obtain dense spectroscopic samples (Fig. \ref{fig:photo}). This could be extremely important for high $z$ samples coming from LSST, Euclid and WFIRST.  The recovery of low $k_\parallel$ modes also opens up a wealth of cross-correlation opportunities with projected fields (e.g.\ lensing) which are restricted to modes transverse to the line of sight.  The measurement of long wavelength fluctuations should also enhance the mapping of the cosmic web at these redshifts, helping to find protoclusters or voids for example \cite{White17}.

In this work we have assumed that we measure the 21-cm field in an unperturbed redshift-space field. In practice, what we observe is the field that has been displaced on large scales by the effect of weak gravitational lensing by the intervening matter along the line of sight. It is clear that the two effects must be degenerate at some level. A location of a given halo in observed redshift-space is obtained by adding the effect of the Lagrangian displacement from the proto-halo region and the effect of weak-lensing by the lower-redshift structures. The latter can be replaced, at least approximately, by a distribution of matter whose Lagrangian displacement absorbs both effects. However, the degeneracy is not perfect. In particular, weak lensing effect is almost exclusively limited to moving the structures in transverse direction by a displacement field that changes slowly with distance. These issues have been studied in the context of traditional lensing estimators in ref.~\cite{1803.04975}, which find encouraging results. Similar lessons should apply to our method, which should, if anything, perform better since it naturally extracts more signal.  We leave this for future investigation.

For the purpose of reconstruction, we have assumed a fixed cosmology and fixed bias parameters. Ideally, one would like to estimate the model and cosmological parameters at the same time as reconstructing linear field. The impact of keeping these parameters fixed at their fiducial values varies with the science objective, for instance it will differently affect BAO reconstruction vs.\ photometric redshift estimation. In some cases, since the non-linear dynamics of our forward model are quite cheap, one can imagine doing this reconstruction recursively, updating model parameters if they vary significantly. A detailed analysis of this is beyond the scope of current work and we leave this for the future. For completeness, we point out that some recent papers (e.g.\ ref.~\cite{Elsner19}) have explored \rsp{ways of taking into account the model
and cosmology parameters simultaneously} for other biased tracers. 

\rsp{We have also made simplifying assumptions to generate our observed mock data to test the reconstruction algorithm.} Since HI assignment to halos is still poorly understood, we have assumed a simple semi-analytic model. We have neglected any effect of UV background fluctuations \cite{Sanderbeck18, HiddenValley19} or assembly bias \cite{VN18} on HI distribution in halos and galaxies. However a flexible bias model, and our forward modeling framework, should be able to include these. In the same vein, it's worth exploring the impact of stochasticity (scatter) between HI mass and dark matter mass on reconstruction since it adds noise on all scales.  However we leave it for future work when the required observations can calibrate the amplitude of the effect.

The success of a quadratic, Lagrangian bias model plus perturbative dynamics (Eq.~\ref{eq:biasmodel1}) in describing the HI field also motivates a new route for making mock catalogs.  As we have demonstrated (Fig.~\ref{fig:biasperf}) our forward model using a function of the linear density and shear field to weight particles which are moved using Zeldovich dynamics (or $2^{\rm nd}$ order perturbation theory) generates a good realization of an HI field in redshift space.  The agreement can be improved even further by dividing by the transfer function at the field level.  Since the initial grid can be relatively coarse (Mpc scale), very large volumes are achievable with reasonable computing resources.  To the HI field generated in this manner one can add light-cone effects, foregrounds, ultra-violet background fluctuations, etc.  While it is beyond the scope of our paper, our results also suggest that such mock catalogs could be used in the modeling of reionization (see also ref.~\cite{McQuinn18}) where the dynamics should be even more linear.  To our model of the HI could be added a simple model for ionization fluctuations (e.g.~those developed in ref.~\cite{Zahn11} or similar).  The model already predicts the velocity field, so the redshift-space clustering of HI on the lightcone or statistics like the kinetic Sunyaev-Zeldovich effect (kSZ; \cite{Sunyaev70}) can be forecast (e.g.~ref.~\cite{Alvarez16}).

While we have proceeded numerically in this paper, Fig.~\ref{fig:biasperf} shows that a Lagrangian bias model with Zeldovich dynamics does quite a good job of describing the low $k$ modes of the HI field.  This suggests it may be possible to develop a fully analytic understanding of our reconstruction process, and the statistics that are being used for clustering redshifts or lensing cross-correlations.  In principle the analytic models could be extended to reionization and kSZ.  We defer development of such models to future work.

\section*{Acknowledgments} 

We would like to thank the Cosmic Visions 21-cm Collaboration for stimulating discussion on the challenges for 21-cm surveys. C.M. would also like to thank Marcel Schmittfull for useful discussions. 
M.W.~is supported by the U.S.~Department of Energy and by NSF grant number 1713791.
This research used resources of the National Energy Research Scientific Computing Center (NERSC), a U.S. Department of Energy Office of Science User Facility operated under Contract No. DE-AC02-05CH11231.
This work made extensive use of the NASA Astrophysics Data System and of the {\tt astro-ph} preprint archive at {\tt arXiv.org}.

\appendix

\section{Validating annealing}
\label{app:validanneal}

\begin{figure}
    \centering
    \resizebox{1\columnwidth}{!}{\includegraphics{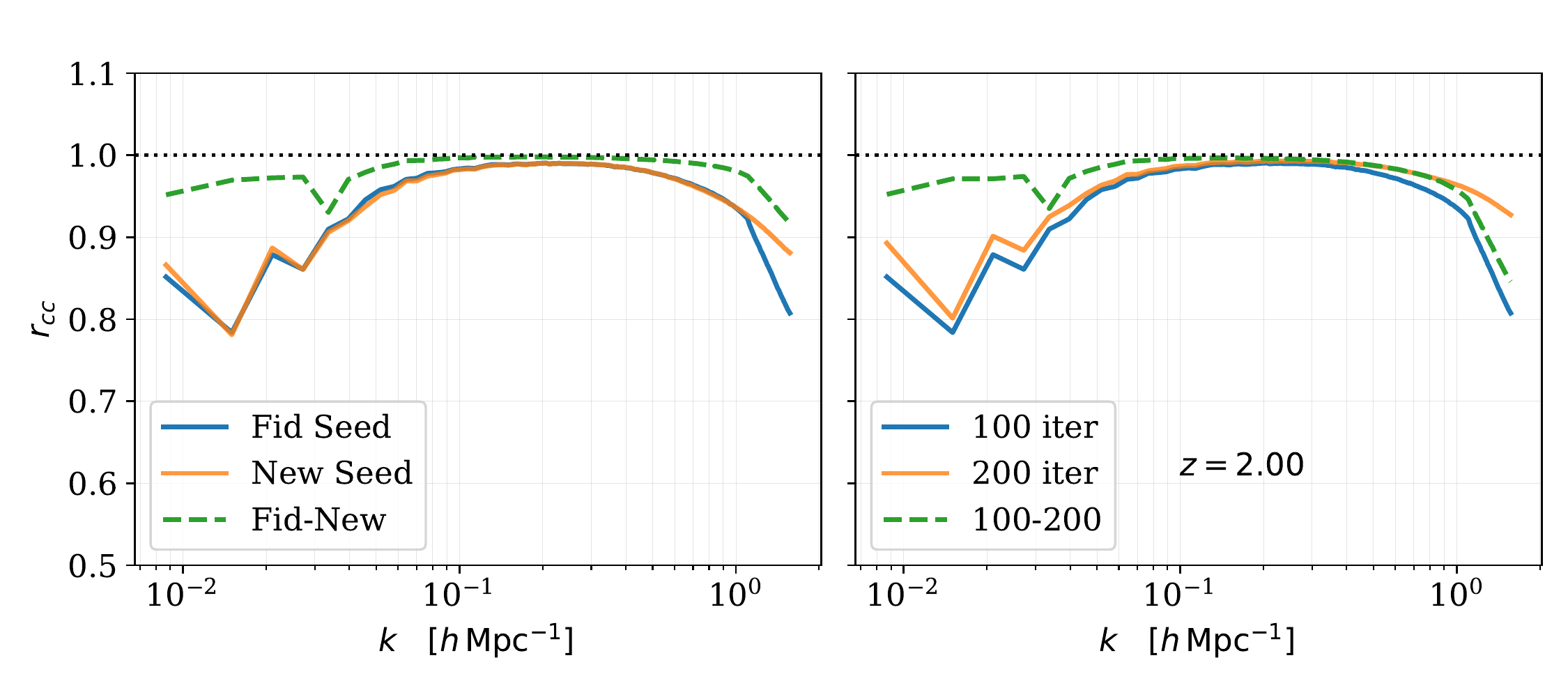}}
    \caption{We show the cross-correlation of the reconstructed data HI field with the true field for different cases. Left: We show reconstructions for different initial conditions (different seed) and show that in both the cases, the reconstructed fields are very well correlated with each other (dashed green), as well as true field (solids). Right: we compare our fiducial annealing scheme (100 iterations) with another reconstruction where we reconstruct for twice the number of steps (200 iterations) and find the improvements to be marginal, as well as the two reconstructed fields correlated well with each other.
    }
\label{fig:anneal}
\end{figure}

Here we show the results of simulations to validate our annealing scheme and convergence of our method. 

Our annealing scheme involves smoothing the residual mesh on different (decreasing) smoothing scales over iterations to reconstruct large scales before smaller scales. The specific choice of smoothing scales does not result in any noticeable difference at the end of convergence as long as we use multiple smoothing scales, starting from large ($>10 \hMpc$) scales. However at every step, we do only 100 iterations, which is motivated by that we do not see the large scales changing significantly. To confirm this, we do a run by increasing the number of iterations and letting the optimizer run for longer.
For every annealing step in this case, we do 200 iterations instead of 100. The result, in terms of cross correlation with the true HI field, is shown in \ref{fig:anneal} and compared with the fiducial 100 iterations. The changes in the cross correlation are primarily on the largest and smallest scales, but less than $3\%$ for scales $k < 0.8 \hMpc$.
Since this marginal change comes at the cost of doubling the computational cost of the reconstruction, for this work we find that our current annealing scheme provides a good balance between the cost and performance but note that the results can further be improved with more computation.

Given the heuristic nature of our annealing scheme, we also validate our convergence to a `correct' solution by running multiple reconstructions starting with different initial conditions. We verify that for all these cases, the differences in the reconstructed fields are small. One example is show in Fig. \ref{fig:anneal}, left plot, for our fiducial seed (for which we show all the results in the text) and a new seed. Both the reconstructed fields are well correlated on all scales of interest with each other ($r_c>0.98$ for $k<1 \hMpc$) and have indistinguishable cross-correlation with the true data field, except on the small scales $k<1 \hMpc$. While this does not guarantee the uniqueness of our solution, a detailed convergence analysis will have to go go hand-in-hand with the uncertainty analysis of the reconstructed field, which is beyond the scope of this work.

\bibliographystyle{JHEP}
\bibliography{main}
\end{document}